\documentclass[english,rmp,twocolumn,aps]{revtex4}
\usepackage[T1]{fontenc}
\usepackage[latin1]{inputenc}
\usepackage{graphicx}
\usepackage{amssymb}

\usepackage{babel}

\begin{document}

\title{The Einstein-Podolsky-Rosen paradox: from concepts to applications }

\author{M. D. Reid}

\affiliation{ARC Centre of Excellence for Quantum-Atom Optics, School of Physical
Sciences, University of Queensland, Brisbane, QLD 4072, Australia }

\author{P. D. Drummond}

\affiliation{ARC Centre of Excellence for Quantum-Atom Optics and Centre for Atom
Optics and Ultrafast Spectroscopy, Swinburne University of Technology,
PO Box 218, Melbourne, VIC 3122 Australia. }

\author{W. P. Bowen}

\affiliation{School of Physical Sciences, University of Queensland, Brisbane,
QLD 4072, Australia}

\author{E. G. Cavalcanti}

\affiliation{Centre for Quantum Dynamics, Griffith University, Brisbane QLD 4111,
Australia}

\author{P. K. Lam, H. A. Bachor}

\affiliation{ARC Centre of Excellence for Quantum-Atom Optics, Building 38, The
Australian National University, Canberra ACT 0200, Australia }

\author{U. L. Andersen}

\affiliation{Department of Physics, Technical University of Denmark, Building
309, 2800 Lyngby, Denmark.}

\author{G. Leuchs}

\affiliation{Institut für Optik, Information and Photonik, Max-Planck Forschungsgruppe,
Universität Erlangen-Nürnberg, Günther-Scharowsky-Str. 1, D-91058
Erlangen, Germany}

\begin{abstract}
This Colloquium examines the field of the EPR \emph{Gedankenexperiment},
from the original paper of Einstein, Podolsky and Rosen, through to
modern theoretical proposals of how to realize both the continuous-variable
and discrete versions of the EPR paradox. We analyze the relationship
with entanglement and Bell's theorem, and summarize the progress to
date towards experimental confirmation of the EPR paradox, with a
detailed treatment of the continuous-variable paradox in laser-based
experiments. Practical techniques covered include continuous-wave
parametric amplifier and optical fibre quantum soliton experiments.
We discuss current proposals for extending EPR experiments to massive-particle
systems, including spin-squeezing, atomic position entanglement, and
quadrature entanglement in ultra-cold atoms. Finally, we examine applications
of this technology to quantum key distribution, quantum teleportation
and entanglement-swapping.

\tableofcontents{} 
\end{abstract}
\maketitle

\section{Introduction}

In 1935, Einstein, Podolsky and Rosen (EPR) originated the famous
{}``EPR paradox'' (\citet{EinsteinPodolskyRosen1935}). This argument
concerns two spatially separated particles which have both perfectly
correlated positions and momenta, as is predicted possible by quantum
mechanics. The EPR paper spurred investigations into the nonlocality
of quantum mechanics, leading to a direct challenge of the philosophies
taken for granted by most physicists. Furthermore, the EPR paradox
brought into sharp focus the concept of entanglement, now considered
to be the underpinning of quantum technology.

Despite its huge significance, relatively little has been done to
directly realize the original EPR \emph{Gedankenexperiment}. Most
published discussion has centred around the testing of theorems by
\citet{Bell1964}, whose work was derived from that of EPR, but proposed
more stringent tests dealing with a different set of measurements.
The purpose of this Colloquium is to give a different perspective.
We go back to EPR's original paper, and analyze the current theoretical
and experimental status, and implications, of the EPR paradox itself:
as an independent body of work.

A paradox is: {}``a seemingly absurd or self-contradictory statement
or proposition that may in fact be true%
\footnote{Compact Oxford English Dictionary, 2006, www.askoxford.com%
}''. The EPR conclusion was based on the assumption of local realism,
and thus the EPR argument pinpoints a contradiction between \emph{local
realism} and the completeness of quantum mechanics. This was therefore
termed a {}``paradox'' by \citet{Schroedinger1935b}, \citet{Bohm1951},
\citet{Bell1964} and \citet{BohmAharonov1957}. EPR took the prevailing
view of their era that \emph{local realism} must be valid. They argued
from this premise that quantum mechanics must be incomplete. With
the insight later provided by \citet{Bell1964}, the EPR argument
is best viewed as the first demonstration of problems arising from
the premise of local realism.

The intention of EPR was to motivate the search for a theory {}``better''
than quantum mechanics. However, EPR never questioned the correctness
of quantum mechanics, only its completeness. They showed that if a
set of assumptions, which we now call local realism, is upheld, then
quantum mechanics must be incomplete. Owing to the subsequent work
of Bell, we now know what EPR didn't know: local realism, the {}``\emph{realistic
philosophy of most working scientists}'' (\citet{ClauserShimony1978}),
is itself in question. Thus, an experimental realization of the EPR
proposal provides a way to demonstrate a type of entanglement inextricably
connected with quantum nonlocality. 

In the sense that the local realistic theory envisaged by them cannot
exist, EPR were {}``wrong''. What EPR did reveal in their paper,
however, was an inconsistency between local realism and the completeness
of quantum mechanics. Hence, we must abandon at least one of these
premises. This was clever, insightful and correct. The EPR paper therefore
provides a way to distinguish quantum mechanics as a complete theory
from classical reality, in a quantitative sense.

The conclusions of the EPR argument can only be drawn if certain correlations
between the positions and momenta of the particles can be confirmed
experimentally. The work of EPR, like that of Bell, requires experimental
demonstration, since it could be supposed that the quantum states
in question are not physically accessible, or that quantum mechanics
itself is wrong. It is not feasible to prepare the perfect correlations
of the original EPR proposal. Instead, we show that the violation
of an inferred Heisenberg Uncertainty Principle -- an {}``EPR inequality''
-- is eminently practical. These EPR inequalities provide a way to
test the incompatibility of local realism, as generalized to a non-deterministic
situation, with the completeness of quantum mechanics. Violating an
EPR inequality is a demonstration of the EPR paradox.

In a nutshell, we will conclude that EPR experiments provide an important
complement to those of Bell. While the conclusions of Bell's theorem
are stronger, the EPR approach is applicable to a greater variety
of physical systems. Most Bell tests have been confined to single
photon counting measurements with discrete outcomes, whereas recent
EPR experiments have involved continuous variable outcomes and high
detection efficiencies. This leads to possibilities for tests of quantum
nonlocality in new regimes involving massive particles and macroscopic
systems. Significantly, new applications in the field of quantum information
are feasible.

In this Colloquium, we outline the theory of EPR's seminal paper,
and also provide an overview of more recent theoretical and experimental
achievements. We discuss the development of the EPR inequalities,
and how they can be applied to quantify the EPR paradox for both spin
and amplitude measurements. A limiting factor for the early spin EPR
experiments of \citet{WuShaknov1950}, \citet{FreedmanClauser1972},
\citet{AspectGrangier1981} and others was the low detection efficiencies,
which meant probabilities were surmised using a postselected ensemble
of counts. In contrast, the more recent EPR experiments report an
amplitude correlation measured over the whole ensemble, to produce
unconditionally, on demand, states that give the entanglement of the
EPR paradox\emph{;} although causal separation was not yet achieved.
We explain in some detail the methodology and development of these
experiments, first performed by \citet{OuPereira1992}.

An experimental realization of the EPR proposal will always imply
entanglement, and we analyze the relationship between entanglement,
the EPR paradox and Bell's theorem. In looking to the future, we review
recent experiments and proposals involving massive particles, ranging
from room-temperature spin-squeezing experiments to proposals for
the EPR-entanglement of quadratures of ultra-cold Bose-Einstein condensates.
A number of possible applications of these novel EPR experiments have
already been proposed, for example in the areas of quantum cryptography
and quantum teleportation. Finally, we discuss these, with emphasis
on those applications that use the form of entanglement closely associated
with the EPR paradox.

\section{The continuous variable EPR paradox}

\citet{EinsteinPodolskyRosen1935} focused attention on the mysteries
of the quantum entangled state by considering the case of two spatially
separated quantum particles that have both maximally correlated momenta
and maximally anti-correlated positions. In their paper entitled \emph{{}``Can
Quantum-Mechanical Description of Physical Reality Be Considered Complete?'',}
they pointed out an apparent inconsistency between such states and
the premise of \emph{local realism}, arguing that this inconsistency
could only be resolved through a completion of quantum mechanics.
Presumably EPR had in mind to supplement quantum theory with a hidden
variable theory, consistent with the {}``elements of reality'' defined
in their paper.

After \citet{Bohm1952} demonstrated that a (non-local) hidden-variable
theory was feasible, subsequent work by \citet{Bell1964} proved the
impossibility of completing quantum mechanics with \emph{local} hidden
variable theories. This resolves the paradox by pointing to a failure
of local realism itself -- at least at the microscopic level. The
EPR argument nevertheless remains significant.

\textbf{It reveals the} \textbf{\emph{necessity}} \textbf{of either
rejecting local realism or completing quantum mechanics (or both).}

\subsection{The 1935 argument: EPR's {}``elements of reality''}

\begin{figure}
\includegraphics[width=7cm]{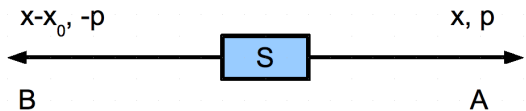}

\caption{(Color online) The original EPR gedanken-experiment. Two particles
move from a source $S$ into spatially separated regions $A$ and
$B$, and yet continue to have maximally correlated positions \emph{and}
anti-correlated momenta. This means one may make an instant prediction,
with $100$\% accuracy, of either the position or momentum of particle
$A$, by performing a measurement at $B$. EPR concluded the results
of both measurements at $A$ pre-exist, in the form of {}``elements
of reality'', and outlined the premises, local realism, rigorously
associated with this reasoning. \label{cap:Schematic-diagram-EPR-original }}

\end{figure}

The EPR argument is based on the premises that are now generally referred
to as \emph{local realism} (quotes are from the original paper): 
\begin{itemize}
\item {}``If, without disturbing a system, we can predict with certainty
the value of a physical quantity'', then {}``there exists an element
of physical reality corresponding to this physical quantity''. The
{}``element of reality'' represents the predetermined value for
the physical quantity. 
\item The locality assumption postulates no action-at-a-distance, so that
measurements at a location $B$ cannot immediately {}``disturb''
the system at a spatially separated location $A$ . 
\end{itemize}
EPR treated the case of a non-factorizable pure state $|\psi\rangle$
which describes the results for measurements performed on two spatially
separated systems at $A$ and $B$ (Fig. \ref{cap:Schematic-diagram-EPR-original }).
{}``Non-factorizable'' means {}``entangled'', that is, we cannot
express $|\psi\rangle$ as a simple product $|\psi\rangle=|\psi\rangle_{A}|\psi\rangle_{B}$,
where $|\psi\rangle_{A}$ and $|\psi\rangle_{B}$ are quantum states
for the results of measurements at $A$ and $B$, respectively.

In the first part of their paper, EPR point out in a general way the
puzzling aspects of such entangled states. The key issue is that one
can expand $|\psi\rangle$ in terms of more than one basis, that correspond
to different experimental settings, which we parametrize by $\phi$.
Consider the state

\begin{equation}
\left|\psi\right\rangle =\int dx\left|\psi_{x}\right\rangle _{\phi,A}\left|u_{x}\right\rangle _{\phi,B}\,.\label{eq:EPRgeneral}\end{equation}
 Here the eigenvalue $x$ could be continuous or discrete. The parameter
setting $\phi$ at the detector $B$ is used to define a particular
orthogonal measurement basis $\left|u_{x}\right\rangle _{\phi,B}$.
On measurement at $B$, this projects out a wave-function $\left|\psi_{x}\right\rangle _{\phi,A}$
at $A$, the process called {}``reduction of the wave packet''.
The puzzling issue is that different choices of measurements $\phi$
at $B$ will cause reduction of the wave packet at A in more than
one possible way. EPR state that, {}``as a consequence of two different
measurements'' at $B$, the {}``second system may be left in states
with two different wavefunctions''. Yet, {}``no real change can
take place in the second system in consequence of anything that may
be done to the first system''.

Despite the apparently acausal nature of state collapse (\citet{Herbert1982}),
the linearity or `nocloning' property of quantum mechanics rules out
superluminal communication (\citet{Diecks1982}; \citet{WoottersZurek1982}).
This clearly supports EPR's original insight. \citet{Schroedinger1935b,Schroedinger1936}
studied this case as well, referring to this apparent influence by
$B$ on the remote system $A$ as {}``steering''.

The problem was crystallized by EPR with a specific example, shown
in Fig. \ref{cap:Schematic-diagram-EPR-original }. EPR considered
two spatially separated subsystems, at $A$ and $B$, each with two
observables $\hat{x}$ and $\hat{p}$ where $\hat{x}$ and $\hat{p}$
are non-commuting quantum operators, with commutator $\left[\hat{x},\hat{p}\right]=\hat{x}\hat{p}-\hat{p}\hat{x}=2C\neq0$.
The results of the measurements $\hat{x}$ and $\hat{p}$ are denoted
$x$ and $p$ respectively, and this convention we follow throughout
the paper. We note that EPR assumed a continuous variable spectrum,
but this is not crucial to the concepts they raised. In our treatment
we will scale the observables so that $C=i$, for simplicity, which
gives rise to the Heisenberg uncertainty relation \begin{equation}
\Delta x\Delta p\ge1\,\,.\label{eq:Heisenberg}\end{equation}
 where $\Delta x$ and $\Delta p$ are the standard deviations in
the results $x$ and $p$, respectively.

EPR considered the quantum wavefunction $\psi$ defined in a position
representation \begin{equation}
\psi\left(x,x^{B}\right)=\int e^{\left(ip/\hbar\right)\left(x-x^{B}-x_{0}\right)}dp\,,\label{eq:EPR-state}\end{equation}
 where $x_{0}$ is a constant implying space-like separation. Here
the pairs $x$ and $p$ refer to the results for position and momentum
measurements at $A$, while $x^{B}$ and $p^{B}$ denote the position
and momentum measurements at $B$. We leave off the superscript for
system $A$, to emphasize the inherent asymmetry that exists in the
EPR argument, where one system $A$ is steered by the other, $B$.

According to quantum mechanics, one can {}``predict with certainty''
that a measurement $\hat{x}$ will give result $x^{B}+x_{0}$, if
a measurement $\hat{x^{B}}$, with result $x^{B}$, was already performed
at $B$. One may also {}``predict with certainty'' the result of
measurement $\hat{p}$, for a different choice of measurement at $B$.
If the momentum at $B$ is measured to be $p$, then the result for
$\hat{p}$ is $-p$. These predictions are made {}``without disturbing
the second system'' at $A$, based on the assumption, implicit in
the original EPR paper, of {}``locality''. The locality assumption
can be strengthened if the measurement events at $A$ and $B$ are
causally separated (such that no signal can travel from one event
to the other, unless faster than the speed of light).

The remainder of the EPR argument may be summarized as follows (\citet{ClauserShimony1978}).
Assuming \emph{local realism}, one deduces that both the measurement
outcomes, for $x$ and $p$ at $A$, are predetermined. The perfect
correlation of $x$ with $x^{B}+x_{0}$ implies the existence of an
{}``element of reality'' for the measurement $\hat{x}$. Similarly,
the correlation of $p$ with $-p^{B}$ implies an {}``element of
reality'' for $\hat{p}$. Although not mentioned by EPR, it will
prove useful to mathematically represent the {}``elements of reality''
for $\hat{x}$ and $\hat{p}$ by the respective variables $\mu_{x}^{A}$
and $\mu_{p}^{A}$ , whose {}``possible values are the predicted
results of the measurement'' (\citet{Mermin1990}).

To continue the argument, \emph{local realism} implies the existence
of two elements of reality, $\mu_{x}^{A}$ and $\mu_{p}^{A}$, that
\emph{simultaneously} predetermine, with \emph{absolute definiteness},
the results for measurement $x$ or $p$ at $A$. These {}``elements
of reality'' for the localized subsystem $A$ are not themselves
consistent with quantum mechanics. Simultaneous determinacy for both
the position and momentum is not possible for any quantum state. Hence,
assuming the validity of local realism, one concludes quantum mechanics
to be incomplete. Bohr's early reply (\citet{Bohr1935}) to EPR was
essentially an intuitive defense of quantum mechanics and a questioning
of the relevance of local realism.

\subsection{Schrödinger's response: entanglement and separability}

It was soon realized that the paradox was intimately related to the
structure of the wavefunction in quantum mechanics, and the opposite
ideas of entanglement and separability. \citet{Schroedinger1935}
pointed out that the EPR two-particle wavefunction in Eq. (\ref{eq:EPR-state})
was \emph{verschränkten} - which he later translated as \emph{entangled}
(\citet{Schroedinger1935b}) - i.e., not of the separable form $\psi_{A}\psi_{B}$.
Both he and \citet{Furry1936} considered as a possible resolution
of the paradox that this {}``entanglement'' degrades as the particles
separate spatially, so that EPR correlations would not be physically
realizable. Experiments considered in this Colloquium show this resolution
to be untenable microscopically, but the proposal led to later theories
which only modify quantum mechanics macroscopically (\citet{GhirardiRimini1986};
\citet{Bell1988}; \citet{BassiGhirardi2003}).

Quantum inseparability (entanglement) for a general mixed quantum
state is defined as the \emph{failure} of \begin{equation}
\widehat{\rho}=\int d{\lambda}P\left(\lambda\right)\widehat{\rho}_{\lambda}^{A}\otimes\widehat{\rho}_{\lambda}^{B}\,\,,\label{eqn:sep}\end{equation}
 where $\int d{\lambda}P(\lambda)=1$ and $\widehat{\rho}$ is the
density operator%
\footnote{In this text, we use {}``entanglement'' in the simplest sense, to
mean a state for a composite system which is nonseparable, so that
(\ref{eqn:sep}) fails. The issues of the EPR paradox that make entanglement
interesting in fact demand that the systems $A$ and $B$ can be spatially
separated, and these are the types of systems we address in this paper.
However, a closer study would also consider restrictions on $A$ and
$B$, for use of the term. This distinction, between a quantum \emph{correlation}
and \emph{entanglement,} is discussed by \citet{Shore2008}.%
}. Here $\lambda$ is a discrete or continuous label for component
states, and $\widehat{\rho}_{\lambda}^{A,B}$ correspond to density
operators that are restricted to the Hilbert spaces $A$,$B$ respectively.

The definition of inseparability extends beyond that of the EPR situation,
in that one considers a whole spectrum of measurement choices, parametrized
by $\theta$ for those performed on system $A$, and by $\phi$ for
those performed on $B$. We introduce the new notation $\hat{x}_{\theta}^{A}$
and $\hat{x}_{\phi}^{B}$ to describe all measurements at $A$ and
$B$. Denoting the eigenstates of $\hat{x}_{\theta}^{A}$ by $|x_{\theta}^{A}\rangle$,
we define $P_{Q}\left(x_{\theta}^{A}\left|\theta,\lambda\right.\right)=\langle x_{\theta}^{A}|\widehat{\rho}_{\lambda}^{A}|x_{\theta}^{A}\rangle$
and $P_{Q}\left(x_{\phi}^{B}\left|\phi,\lambda\right.\right)=\langle x_{\phi}^{B}|\widehat{\rho}_{\lambda}^{B}|x_{\phi}^{B}\rangle$,
which are the localized probabilities for observing results $x_{\theta}^{A}$
and $x_{\phi}^{B}$ respectively. The separability condition (\ref{eqn:sep})
then implies that joint probabilities $P(x_{\theta}^{A},x_{\phi}^{B})$
are given as: \textbf{\begin{equation}
P\left(x_{\theta}^{A},x_{\phi}^{B}\right)=\int d\lambda P\left(\lambda\right)P_{Q}\left(x_{\theta}^{A}\left|\lambda\right.\right)P_{Q}\left(x_{\phi}^{B}\left|\lambda\right.\right)\,\,.\label{eqn:sepprob}\end{equation}
}  We note the restriction, that for example $\Delta^{2}(x^{A}|\lambda)\Delta^{2}(p^{A}|\lambda)\geq1$
where $\Delta^{2}(x^{A}|\lambda)$ and $\Delta^{2}(p^{A}|\lambda)$
are the variances of $P_{Q}\left(x_{\theta}^{A}\left|\theta,\lambda\right.\right)$
for the choices $\theta$ corresponding to position $x$ and momentum
$p$, respectively. The original EPR state of Eq. (\ref{eq:EPR-state})
is not separable.

The most precise signatures of entanglement rely on entropic or more
general information-theoretic measures. This can be seen in its simplest
form when $\widehat{\rho}$ is a pure state, so that $Tr\widehat{\rho}^{2}=1$.
Under these conditions, it follows that $\widehat{\rho}$ is entangled
if and only if the von Neumann entropy measure of either reduced density
matrix $\widehat{\rho}^{A}=Tr_{B}\widehat{\rho}$ or $\widehat{\rho}^{B}=Tr_{A}\widehat{\rho}$
is positive. Here the entropy is defined as:\begin{equation}
S[\widehat{\rho}]=-Tr\widehat{\rho}\ln\widehat{\rho}\end{equation}
 When $\widehat{\rho}$ is a mixed state, one must turn to variational
measures like the entanglement of formation to obtain necessary and
sufficient measures (\citet{BennettDiVincenzo1996}). The entanglement
of formation leads to the popular concurrence measure for two qubits
(\citet{Wootters1998}). A necessary but not sufficient measure for
entanglement is the partial transpose criterion of \citet{Peres1996}.

\section{Discrete spin variables and Bell's theorem}

\subsection{The EPR-Bohm paradox: early EPR experiments}

\begin{figure}
\includegraphics[width=7cm]{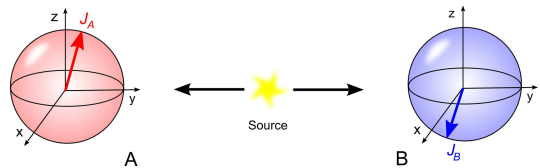}

\caption{(Color online) The Bohm gedanken EPR experiment. Two spin-$\frac{1}{2}$
particles prepared in a singlet state move from the source into spatially
separated regions $A$ and $B$, and give anti-correlated outcomes
for $J_{\theta}^{A}$ and $J_{\theta}^{B}$, $ $where $\theta$ is
$x$, $y$ or $z$. \label{cap:Schematic-diagram-Bohm-original}}

\end{figure}

As the continuous-variable EPR proposal was not experimentally realizable
at the time, much of the early work relied on an adaptation of the
EPR paradox to spin measurements by \citet{Bohm1951}, as depicted
in Fig. (\ref{cap:Schematic-diagram-Bohm-original}).

This corresponds to the general form given in Eq. (\ref{eq:EPRgeneral}).
Specifically, Bohm considered two spatially-separated spin-$1/2$
particles at $A$ and $B$ produced in an entangled singlet state
(often referred to as the {}``EPR-Bohm state'' or the {}``Bell-state''):

\begin{equation}
\left|\psi\right\rangle =\frac{1}{\sqrt{2}}\left(\left|\frac{1}{2}\right\rangle _{A}\left|-\frac{1}{2}\right\rangle _{B}-\left|-\frac{1}{2}\right\rangle _{A}\left|\frac{1}{2}\right\rangle _{B}\right)\label{eq:bellstate}\end{equation}
 Here $|\pm\frac{1}{2}\rangle_{A}$ are eigenstates of the spin operator
$\widehat{J}_{z}^{A}$, and we use $\widehat{J}_{z}^{A}$ , $\widehat{J}_{x}^{A}$
, $\widehat{J}_{y}^{A}$ to define the spin-components measured at
location $A$. The spin-eigenstates and measurements at $B$ are defined
similarly. By considering different quantization axes, one obtains
different but equivalent expansions of $\left|\psi\right\rangle $
in Eq. (\ref{eq:EPRgeneral}), just as EPR suggested.

Bohm's reasoning is based on the existence, for Eq. (\ref{eq:bellstate}),
of a maximum anti-correlation between not only $\widehat{J}_{z}^{A}$
and $\widehat{J}_{z}^{B}$, but $\widehat{J}_{y}^{A}$ and $\widehat{J}_{y}^{B}$,
and also $\widehat{J}_{x}^{A}$ and $\widehat{J}_{x}^{B}$. An assumption
of local realism would lead to the conclusion that the three spin
components of particle $A$ were simultaneously predetermined, with
absolute definiteness. Since no such quantum description exists, this
is the situation of an EPR paradox. A simple explanation of the discrete-variable
EPR paradox has been presented by \citet{Mermin1990} in relation
to the three-particle Greenberger-Horne-Zeilinger correlation (\citet{GreenbergerHorne1989}).

An early attempt to realize EPR-Bohm correlations for discrete (spin)
variables came from \citet{BleulerBradt1948}, who examined the gamma-radiation
emitted from positron annihilation. These are spin-one particles which
form an entangled singlet. Here, correlations were measured between
the polarizations of emitted photons, but with very inefficient Compton-scattering
polarizers and detectors, and no control of causal separation. Several
further experiments were performed along similar lines (\citet{WuShaknov1950}),
as well as with correlated protons (\citet{Lamehi-RachtiMittig1976}).
While these are sometimes regarded as demonstrating the EPR paradox
(\citet{BohmAharonov1957}), the fact that they involved extremely
inefficient detectors, with postselection of coincidence counts, makes
this interpretation debatable.

\subsection{Bell's theorem}

The EPR paper concludes by referring to theories that might complete
quantum mechanics: {}``..\emph{we have left open the question of
whether or not such a description exists. We believe, however, that
such a theory is possible}''. The seminal works of \citet{Bell1964,Bell1988}
and \citet{ClauserHorne1969} (CHSH) clarified this issue, to show
that this speculation was wrong\textbf{.} Bell showed that the predictions
of \emph{local hidden variable theories} (LHV) differ from those of
quantum mechanics, for the {}``Bell state'', Eq. (\ref{eq:bellstate}).

Bell-CHSH considered theories for two spatially-separated subsystems
$A$ and $B$. As with separable states, Eq. (\ref{eqn:sep}) and
Eq. (\ref{eqn:sepprob}), it is assumed there exist parameters $\lambda$
that are shared between the subsystems and which denote localized
-- though not necessarily quantum -- states for each. Measurements
can be performed on $A$ and $B$, and the measurement choice is parametrized
by $\theta$ and $\phi$, respectively. Thus for example, $\theta$
may be chosen to be either position and momentum, as in the original
EPR gedanken experiment, or an analyzer angle as in the Bohm-EPR gedanken
experiment. We denote the result of the measurement labelled $\theta$
at $A$ as $x_{\theta}^{A}$, and use similar notation for outcomes
at $B$. The assumption of Bell's \emph{locality} is that the probability
$P\left(x_{\theta}^{A}\left|\lambda\right.\right)$ for $x_{\theta}^{A}$
depends on $\lambda$ and $\theta$, but is independent of $\phi$;
and similarly for $P\left(x_{\phi}^{B}\left|\lambda\right.\right)$.
The {}``local hidden variable'' assumption of Bell and CHSH then
implies the joint probability $P(x_{\theta}^{A},x_{\phi}^{B})$ to
be

\begin{eqnarray}
P\left(x_{\theta}^{A},x_{\phi}^{B}\right) & = & \int_{\lambda}d\lambda P(\lambda)P\left(x_{\theta}^{A}\left|\lambda\right.\right)P\left(x_{\phi}^{B}\left|\lambda\right.\right),\label{eqn:bellsep1}\end{eqnarray}
 where $P(\lambda)$ is the distribution for the $\lambda$. This
assumption, which we call {}``\emph{Bell-CHSH local realism}'',
differs from Eq. (\ref{eqn:sepprob}) for separability, in that the
probabilities $P(x_{\theta}^{A}|\lambda)$ and $P(x_{\phi}^{B}|\lambda)$
do not arise from localised quantum states. From the assumption Eq.
(\ref{eqn:bellsep1}) of LHV, Bell and CHSH derived constraints, famously
referred to as Bell's inequalities. They showed that quantum mechanics
predicts a violation for efficient measurements made on Bohm's entangled
state, Eq. (\ref{eq:bellstate}).

Bell's work provided a resolution of the EPR paradox, in the sense
that a measured violation would indicate a failure of local realism.
While Bell's assumption of local hidden variables is not formally
identical to that of EPR's local realism, one can be extrapolated
from the other (Section VI.A.3). The failure of local hidden variables
is then indicative of a failure of local realism.

\subsection{Experimental tests of Bell's theorem}

A violation of modified Bell inequalities, that employ auxiliary fair-sampling
assumptions (\citet{ClauserShimony1978}), has been achieved by \citet{FreedmanClauser1972},
\citet{KasdayUllman1970}, \citet{FryThompson1976}, \citet{AspectGrangier1981},
\citet{ShihAlley1988}, \citet{OuMandel1988} and others. Most of
these experiments employ photon pairs created via atomic transitions
or using non-linear optical techniques such as optical parametric
amplification. These methods provide an exquisite source of highly
entangled photons in a Bell-state. Causal separation was achieved
by \citet{AspectDalibard1982}, with subsequent improvements by \citet{WeihsJennewein1998}.

However, the low optical and photo-detector efficiencies for counting
individual photons ($\sim5\%$ in the \citet{WeihsJennewein1998}
experiment) prevent the original Bell inequality from being violated.
The original Bell inequality requires a threshold efficiency of $83\%$
($\eta\sim0.83)$ per detector (\citet{GargMermin1987}; \citet{ClauserShimony1978};
\citet{FryWalther1995}), in order to exclude all local hidden variable
theories. For lower efficiencies, one can construct local hidden variable
theories to explain the observed correlations (\citet{ClauserHorne1974};
\citet{Larsson1999}). Nevertheless, these experiments, elegantly
summarized by \citet{Zeilinger1999} and \citet{Aspect2002}, exclude
the most appealing local realistic theories and thus represent strong
evidence in favor of abandoning the local realism premise.

While highly efficient experimental violations of Bell's inequalities
in ion traps (\citet{RoweKielpinski2001}) have been reported, these
have been limited to situations of poor spatial separation between
measurements on subsystems. A conclusive experiment would require
both high efficiency \emph{and} causal separations, as suggested by
\citet{KwiatEberhard1994}, and \citet{FryWalther1995}. Reported
system efficiencies are currently up to $51\%$ (\citet{UrenSilberhorne2004}),
while typical photo-diode single-photon detection efficiencies are
now $60\%$ or more (\citet{PolyakovMigdal2007}), and further improvements
up to $88\%$ with more specialized detectors (\citet{TakeuchiKim1999})
makes a future loophole-free experiment not impossible.

\section{EPR argument for real particles and fields }

In this Colloquium, we focus on the realization of the original EPR
paradox. To recreate the precise gedanken proposal of EPR, one needs
perfect correlations between the positions of two separated particles,
and also between their momenta. This is physically impossible, in
practice.

In order to demonstrate the existence of EPR correlations for real
experiments, one therefore needs to minimally extend the EPR argument,
in particular their definition of local realism, to situations where
there is less than perfect correlation%
\footnote{The extension of local realism, to allow for real experiments, was
also necessary in the Bell case (\citet{ClauserShimony1978}). Bell's
original inequality (\citet{Bell1964}) pertained only to local hidden
variables that predetermine outcomes of spin with absolute certainty.
These deterministic hidden variables follow naturally from EPR's local
realism in a situation of \emph{perfect} correlation, but were too
restrictive otherwise. Further Bell and CHSH inequalities (\citet{ClauserHorne1969};
\citet{Bell1971}; \citet{ClauserHorne1974}) were derived that allow
for a \emph{stochastic} predeterminism, where local hidden variables
give \emph{probabilistic} predictions for measurements. This stochastic
local realism of Bell-CHSH follows naturally from the stochastic extension
of EPR's local realism to be given here, as explained in Section VI.A.\label{fn:lr-1}%
}. We point out that near perfect correlation of the detected photon
pairs has been achieved in the seminal {}``a posteriori'' realization
of the EPR gedanken experiment by \citet{AspectGrangier1981}. However,
it is debatable whether this can be regarded as a rigorous EPR experiment,
because for the full ensemble, most counts at one detector correspond
to no detection at the other.

The stochastic extension of EPR's \emph{local realism} is that one
can predict with a \emph{specified probability distribution} repeated
outcomes of a measurement, remotely, so the {}``values'' of the
elements of reality are in fact those probability distributions. This
definition is the meaning of {}``local realism'' in the text below.
As considered by \citet{Furry1936} and \citet{Reid1989}, this allows
the derivation of an inequality whose violation indicates the EPR
paradox.

We consider non-commuting observables associated with a subsystem
at $A$, in the realistic case where measurements made at $B$ do
not allow the prediction of outcomes at $A$ to be made with certainty.
Like EPR, we assume causal separation of the observations and the
validity of quantum mechanics. Our approach applies to any non-commuting
observables, and we focus in turn on the continuous variable and discrete
cases.

\subsection{Inferred Heisenberg inequality: continuous variable case}

Suppose that, based on a result $x^{B}$ for the measurement at $B$,
an estimate $x_{est}\left(x_{B}\right)$ is made of the result $x$
at $A$. We may define the average error $\Delta_{inf}x$ of this
inference as the root mean square (RMS) of the deviation of the estimate
from the actual value, so that \begin{equation}
\Delta_{inf}^{2}x=\int dxdx^{B}P\left(x,x_{B}\right)\left(x-x_{est}\left(x_{B}\right)\right)^{2}\,\,.\label{eq:infvarx}\end{equation}
 An inference variance $\Delta_{inf}^{2}p$ is defined similarly.

The best estimate, which minimizes $\Delta_{inf}x$, is given by choosing
$x_{est}$ for each $x^{B}$ to be the mean $\langle x|x^{B}\rangle$
of the conditional distribution $P\left(x\left|x^{B}\right.\right)$.
This is seen upon noting that for \emph{each} result $x^{B}$, we
can define the RMS error in each estimate as\begin{equation}
\Delta_{inf}^{2}\left(x\left|x^{B}\right.\right)\,=\int dxP\left(x\left|x^{B}\right.\right)\left(x-x_{est}\left(x^{B}\right)\right)^{2}\,\,.\end{equation}
 The average error in each inference is minimized for $x_{est}=\langle x|x^{B}\rangle\,$,
when each $\Delta_{inf}^{2}\left(x\left|x^{B}\right.\right)\,$ becomes
the variance $\Delta^{2}(x|x^{B})$ of $P\left(x\left|x^{B}\right.\right)$.

We thus define the \emph{minimum} inference error $\Delta_{inf}x$
for position, averaged over all possible values of $x^{B}$, as\begin{eqnarray}
V_{A|B}^{x} & = & \Delta_{inf}^{2}x\Bigl|_{min}=\int dx^{B}P\left(x^{B}\right)\Delta^{2}\left(x\left|x^{B}\right.\right)\,\,,\label{eq:infx}\end{eqnarray}
 where $P\left(x^{B}\right)$ is the probability for a result $x^{B}$
upon measurement of $\hat{x}^{B}.$ This minimized inference variance
is the average of the individual variances for each outcome at $B$.
Similarly, we can define a minimum inference variance, $V_{A|B}^{p}$
, for momentum.

We now derive the EPR criterion applicable to this more general situation.
We follow the logic of the original argument, as outlined in Section
II. Referring back to Fig. (\ref{cap:Schematic-diagram-EPR-original }),
we remember that if we assume local realism, there will exist a predetermination
of the results for both $x$ and $p$. In this case, however, the
predetermination is probabilistic, because we cannot {}``predict
with certainty'' the result $x$. We can predict the \emph{probability}
for $x$ however, based on remote measurement at $B$. We recall the
{}``element of reality'' is a variable, ascribed to the local system
$A$, as part of a theory, to quantify this predetermination. The
{}``element of reality'' $\mu_{x}^{A}$ associated with $\hat{x}$
is, in the words of \citet{Mermin1990} that \textbf{{}``}\emph{predictable
value}'' for a measurement at $A$, based on a measurement at $B$,
which {}``ought to exist whether or not we actually carry out the
procedure necessary for its prediction, since this in no way disturbs
it''. Given the EPR premise and our extension of it, we deduce that
{}``elements of reality'' still exist, but the {}``predictable
values'' associated with them are now probability distributions.

This requires an extension to the definition of the element of reality.
As before, the $\mu_{x}^{A}$ is a variable which takes on certain
values, but the values no longer represent a single predicted outcome
for result $x$ at $A$, but rather they represent a \emph{predicted
probability distribution} for the results $x$ at $A$. Thus each
value for $\mu_{x}^{A}$ defines a probability distribution for $x$.
Since the set of predicted distributions are the conditionals $P(x|x^{B})$,
one for each value of $x^{B}$, the logical choice is to label the
element of reality by the outcomes $x^{B}$, but bearing in mind the
set of predetermined results is \emph{not} the set $\bigl\{ x^{B}\bigr\}$,
but is the set of associated conditional distributions $\bigl\{ P(x|x^{B})\bigr\}$.
Thus we say if the element of reality $\mu_{x}^{A}$ takes the value
$x^{B}$, then the predicted outcome for $x$ is given probabilistically
as $P(x|x^{B})$.

Such probability distributions are also implicit in the extensions
by \citet{ClauserHorne1969} and \citet{Bell1988} of Bell's theorem
to systems of less-than-ideal correlation. The $P(x_{\theta}^{A}|\lambda)$
used in Eq. (\ref{eqn:bellsep1}) is the probability for a result
at $A$ given a hidden variable $\lambda$. The {}``element of reality''
and {}``hidden variable'' have similar meanings, except that the
element of reality is a special {}``hidden variable'' following
from the EPR logic.

To recap the argument, we define $\mu_{x}^{A}$ as a variable whose
values, mathematically speaking, are the set of possible outcomes
$x^{B}$. We also define $P(x|\mu_{x}^{A})$ as the probability of
observing the value $x$ for the measurement $\hat{x}$, in a system
$A$ specified by the `element of reality' $\mu_{x}^{A}$. We might
also ask, what is the probability that the element of reality has
a certain value, namely, what is $P(\mu_{x}^{A})$? Clearly, a particular
value for $\mu_{x}^{A}$ occurs with probability $P\left(\mu_{x}^{A}\right)=P(x^{B})$.
This is because in the local realism framework, the action of measurement
at $B$ (to get outcome $x^{B}$) cannot create the value of the element
of reality $\mu_{x}^{A}$, yet it informs us of its value.

An analogous reasoning will imply probabilistic elements of reality
for $p$ at $A$, with the result that two elements of reality $\mu_{x}^{A},\mu_{p}^{A}$
are introduced to \emph{simultaneously} describe results for the localized
system $A$. We introduce a joint probability distribution $P(\mu_{x}^{A},\mu_{p}^{A})$
for the values assumed by these elements of reality.

It is straightforward to show from the definition of Eq (\ref{eq:infx})
that if $V_{A|B}^{x}V_{A|B}^{p}<1$, then the pair of elements of
reality for $A$ cannot be consistent with a quantum wave-function.
This indicates an inconsistency of \emph{local realism} with the \emph{completeness
of quantum mechanics.} To do this, we quantify the statistical properties
of the elements of reality by defining $\Delta^{2}\left(x|\mu_{x}^{A}\right)$
and $\Delta^{2}\left(p|{\mu}_{p}^{A}\right)$ as the variances of
the probability distributions $P(x|\mu_{x}^{A})$ and $P(p|\mu_{p}^{A})$.
Thus the measurable inference variance is a measure of the average
indeterminacy: \begin{eqnarray}
V_{A|B}^{x} & = & \int d{\mu_{x}^{A}}P(\mu_{x}^{A})\Delta^{2}\left(x|\mu_{x}^{A}\right)\label{eq:veprinfel}\\
 & = & \int d\mu_{x}^{A}d\mu_{p}^{A}P(\mu_{x}^{A},\mu_{p}^{A})\Delta^{2}\left(x|\mu_{x}^{A}\right)\nonumber \end{eqnarray}
 (similarly for $V_{A|B}^{p}$ and $\Delta_{inf}^{2}p$). The assumption
that the state depicted by a particular pair $\mu_{x}^{A}$, $\mu_{p}^{A}$
has an equivalent \emph{quantum} description demands that the conditional
probabilities satisfy the same relations as the probabilities for
a quantum state. For example, if $x$ and $p$ satisfy $\Delta x\Delta p\geq1$,
then $\Delta\left(x|\mu_{x}^{A}\right)\Delta\left(p|{\mu}_{p}^{A}\right)\geq1$\emph{.}
Simple application of the Cauchy-Schwarz inequality gives\begin{eqnarray}
\Delta_{inf}x\Delta_{inf}p & \geq & V_{A|B}^{x}V_{A|B}^{p}\label{eq:proofcs}\\
 & = & \langle\Delta^{2}\left(x|\mu_{x}^{A}\right)\rangle\langle\Delta^{2}\left(p|\mu_{p}^{A}\right)\rangle\nonumber \\
 & \geq & |\langle\Delta\left(x|\mu_{x}^{A}\right)\Delta\left(p|\mu_{p}^{A}\right)\rangle|^{2}\geq1\nonumber \end{eqnarray}
 Thus the observation of $V_{A|B}^{x}V_{A|B}^{p}<1$, or more generally,
\begin{equation}
\Delta_{inf}x\Delta_{inf}p<1\label{eq:infcasch}\end{equation}
 is an EPR criterion, meaning that this would imply an EPR paradox
(\citet{Reid1989,Reid2004}).

One can in principle use any quantum uncertainty constraint (\citet{CavalcantiReid2007}).
Take for example, the relation $\Delta^{2}\left(x|\mu_{x}^{A}\right)+\Delta^{2}\left(p|{\mu}_{p}^{A}\right)\geq2$,
which follows from that of Heisenberg. From this we derive $V_{A|B}^{x}+V_{A|B}^{p}\geq2$,
to imply that \begin{equation}
\Delta_{inf}^{2}x+\Delta_{inf}^{2}p<2\label{eq:duanhalf}\end{equation}
 is also an EPR criterion. On the face of it, this is less useful;
since if (\ref{eq:duanhalf}) holds, then (\ref{eq:infcasch}) must
also hold.

\subsection{Criteria for the discrete EPR paradox}

The discrete variant of the EPR paradox was treated in Section III.
Conclusive experimental realization of this paradox needs to account
for imperfect sources and detectors, just as in the continuous variable
case.

Criteria sufficient to demonstrate Bohm's EPR paradox can be derived
with the inferred uncertainty approach. Using the Heisenberg spin
uncertainty relation \textbf{$\Delta J_{x}^{A}\Delta J_{y}^{A}\geq\left|\left\langle J_{z}^{A}\right\rangle \right|/2$,}
one obtains (\citet{CavalcantiReid2007}) the following spin-EPR criterion
that is useful for the Bell state Eq. (\ref{eq:bellstate}): \begin{eqnarray}
\Delta_{inf}J_{x}^{A}\Delta_{inf}J_{y}^{A} & < & \frac{1}{2}\sum_{{J_{z}^{B}}}P\left(J_{z}^{B}\right)\left|\left\langle J_{z}^{A}\right\rangle _{J_{z}^{B}}\right|\,\,.\label{eq:bohmcrit}\end{eqnarray}
 Here $\left\langle J_{z}^{A}\right\rangle _{J_{z}^{B}}$ is the mean
of the conditional distribution $P\left(J_{z}^{A}|J_{z}^{B}\right)$.
Calculations for Eq. (\ref{eq:bellstate}) including the effect of
detection efficiency $\eta$ reveals this EPR criterion to be satisfied
for $\eta>0.62$. Further spin-EPR inequalities have recently been
derived (\citet{CavalcantiDrummond2007}), employing quantum uncertainty
relations involving sums, rather than the products (\citet{HofmannTakeuchi2003}).
A constraint on the degree of mixing that can still permit an EPR
paradox for the Bell state of Eq. (\ref{eq:bellstate}) can be deduced
from an analysis by \citet{WisemanJones2007}. These authors report
that the \citet{Werner1989} state $\rho_{w}=(1-p_{W})\frac{\mathbb{I}}{4}+p_{W}|\psi\left\rangle \right\langle \psi|$,
which is a mixed Bell state, requires $p_{W}>0.5$ to demonstrate
{}``steering'', which we show in Section VI.A is a necessary condition
for the EPR paradox.

The concept of spin-EPR has been experimentally tested \emph{in the
continuum limit} with purely optical systems for states where $\left\langle J_{z}^{A}\right\rangle \neq0$.
In this case the EPR criterion, linked closely to a definition of
spin squeezing (\citet{KitagawaUeda1993}; \citet{SorensonDuan2001};
\citet{KorolkovaLeuchs2002}; \citet{BowenSchnabel2002}), \begin{equation}
\Delta_{inf}J_{x}^{A}\Delta_{inf}J_{y}^{A}<\frac{1}{2}\left|\left\langle J_{z}^{A}\right\rangle \right|\label{eq:spincrit}\end{equation}
 has been derived by \citet{BowenTreps2002}, and used to demonstrate
the EPR paradox, as summarized in Section VII. Here the correlation
is described in terms of Stokes operators for the polarization of
the fields. The experiments take the limit of large spin values to
make a continuum of outcomes, so high efficiency detectors are used.

We can now turn to the question of whether existing spin-half or two-photon
experiments were able to conclusively demonstrate an EPR paradox.
This depends on the overall efficiency, as in the Bell inequality
case. Generating and detecting pairs of photons is generally rather
inefficient, although results of up to $51\%$ were reported by \citet{UrenSilberhorne2004}.
This is lower than the $62\%$ threshold given above. We conclude
that efficiencies for these types of discrete experiment are still
too low, although there have been steady improvements. The required
level appears feasible as optical technologies improve.

\subsection{A practical linear-estimate criterion for EPR}

\label{linearestimate}

It is not always easy to measure conditional distributions. Nevertheless,
an inference variance, which is the variance of the conditional distribution,
has been so measured for twin beam intensity distributions by \citet{ZhangKasai2003},
who achieved $\Delta_{inf}^{2}$x=0.62.

It is also possible to demonstrate an EPR correlation using criteria
based on the measurement of a sufficiently reduced noise in the appropriate
sum or difference $x-gx^{B}$ and $p+g'p^{B}$ (where here $g$, $g'$
are real numbers). This was proposed by \citet{Reid1989} as a practical
procedure for measuring EPR correlations.

Suppose that an estimate $x_{est}$ of the result for $\hat{x}$ at
$A$, based on a result $x^{B}$ for measurement at $B$, is of the
linear form $x_{est}=gx^{B}+d$. The best linear estimate $x_{est}$
is the one that will minimize \begin{equation}
\Delta_{inf}^{2}x=\left\langle \left\{ x-\left(gx^{B}+d\right)\right\} ^{2}\right\rangle \label{eq:delinfg}\end{equation}
 The best choices for $g$ and $d$ minimize $\Delta_{inf}^{2}x$
and can be adjusted by experiment, or calculated by linear regression
to be $d=\left\langle x-gx^{B}\right\rangle $, $g=\left\langle x,x^{B}\right\rangle /\Delta^{2}x^{B}$
(where we define $\left\langle x,x^{B}\right\rangle =\left\langle xx^{B}\right\rangle -\left\langle x\right\rangle \left\langle x^{B}\right\rangle \,$).
There is also an analogous optimum for the value of $g'$. This gives
a predicted minimum (for linear estimates) of\begin{equation}
\Delta_{inf}^{2}x\mid_{min,L}=\Delta^{2}\left(x-gx^{B}\right)=\Delta^{2}x-\frac{\langle x,x^{B}\rangle^{2}}{\Delta^{2}x^{B}}\label{eq:losepr}\end{equation}
 We note that for Gaussian states (Section VI) this best linear estimate
for $x$, given $x^{B}$, is equal to the mean of the conditional
distribution $P(x|x^{B})$, so that $\Delta_{inf}^{2}x\Bigl|_{min,L}=V{}_{A|B}^{x}$
where $V_{A|B}^{x}$ is the variance of the conditional distribution,
and this approach thus automatically gives the minimum possible $\Delta_{inf}x$.

The observation of \begin{equation}
\Delta^{2}\left(x-gx^{B}\right)\Delta^{2}\left(p+g'p^{B}\right)<1\label{eq:1989crit}\end{equation}
 is sufficient to imply Eq. (\ref{eqn:eprcritt}), which is the condition
for the correlation of the original EPR paradox. This was first experimentally
achieved by \citet{OuPereira1992}.

We note it is also possible to present an EPR criterion in terms of
the sum of the variances. Using (\ref{eq:duanhalf}), on putting $\Delta_{inf}^{2}x=\Delta^{2}\left(x-gx^{B}\right)$
and $\Delta_{inf}^{2}p=\Delta^{2}\left(p+g'p^{B}\right)$ we arrive
at the linear EPR criterion \begin{equation}
\Delta^{2}(x-gx^{B})+\Delta^{2}(p+g'p^{B})<2.\label{eq:duaneprhalfI}\end{equation}

Strictly speaking, to carry out a true EPR gedanken experiment, one
must measure, preferably with causal separation, the separate \emph{values}
for the EPR observables $x$, $x^{B}$, $p$ and $p^{B}$.

\subsection{Experimental criteria for demonstrating the paradox}

We now summarize experimental criteria sufficient to realize the EPR
paradox. To achieve this, one must have two spatially separated subsystems
at $A$ and $B$.

(1): First, to realize the EPR paradox in the spirit intended by EPR
it is necessary that measurement events at $A$ and $B$ be \textbf{\emph{causally
separated}}. This point has been extensively discussed in literature
on Bell's inequalities and is needed to justify the locality assumption,
given that EPR assumed idealized instantaneous measurements. If $c$
is the speed of light and $t_{A}$ and $t_{B}$ are the times of flight
from the source to A and B, then the measurement duration $\Delta t$,
time for the measurements at $A$ and $B$ and the separation $L$
between the subsystems must satisfy\begin{equation}
L>c(t_{A}-t_{B}+\Delta t).\label{eqn:eprcas}\end{equation}

(2): Second, one establishes a \textbf{\emph{prediction protocol}},
so that for each possible outcome of a measurement at $B$, one can
make a prediction about the outcome at $A$. There must be a \textbf{\emph{sufficient
correlation}} between measurements made at $A$ and $B$. The EPR
correlation is demonstrated when the product of the average errors
in the inferred results $x_{est}$ and $p_{est}$ for $\hat{x}$ and
$\hat{p}$ at $A$ falls below a bound determined by the corresponding
Heisenberg Uncertainty Principle.

In the continuous variable case where $x$ and $p$ are such that
$\Delta x\Delta p\geq1$ this amounts to\begin{equation}
\mathcal{E}=\Delta_{inf}x\Delta_{inf}p<1,\label{eqn:eprcritt}\end{equation}
 where we introduce for use in later sections a symbol $\mathcal{E}$
for the measure of the inference (conditional variance) product \textbf{\emph{$\Delta_{inf}x\Delta_{inf}p$.}}
Similar criteria hold for discrete spin variables.

\section{Theoretical model for a continuous variable EPR Experiment}

\subsection{Two-mode squeezed states}

As a physically realizable example of the original continuous variable
EPR proposal, suppose the two systems $A$ and $B$ are localized
modes of the electromagnetic field, with frequencies $\omega_{A,B}$
and boson operators $\widehat{a}$ and $\widehat{b}$ respectively.
These can be prepared in an EPR-correlated state using parametric
down conversion (\citet{ReidDrummond1988,ReidDrummond1989,DrummondReid1990}).
Using a coherent pump laser at frequency $\omega_{A}+\omega_{B}$,
and a nonlinear optical crystal which is phase-matched at these wavelengths,
energy is transferred to the modes. As a result, these modes become
correlated.

\begin{figure}
\includegraphics[width=7cm]{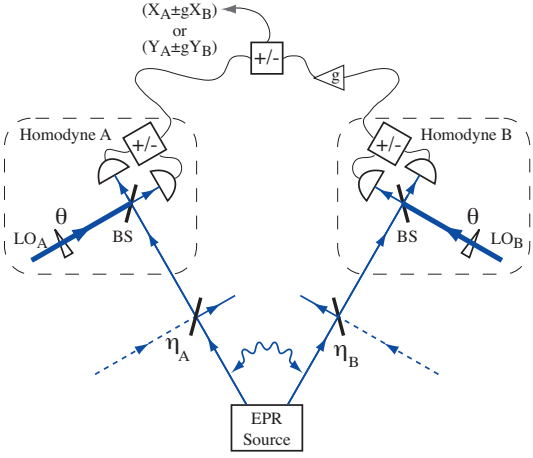}

\caption{(Color online) Schematic diagram of the measurement of the EPR paradox
using field quadrature phase amplitudes. Spatially separated fields
$A$ and $B$ radiate outwards from the EPR source, usually Eq. (\ref{eqn:twomode}).
The field quadrature amplitudes are symbolised $Y$ and $X$. The
fields combine with an intense local oscillator $LO$ field, at beam
splitters $BS$. The outputs of each BS are detected by photodiodes
and their difference current is proportional to the amplitude $Y$
or $X$, depending on the phase shift $\theta$. A gain $g$ is introduced
to read out the final conditional variances, Eq. (\ref{eq:condvarhd}).
Here $\eta_{A}$ and $\eta_{B}$ are the non-ideal efficiencies that
model losses, defined in Section V. \label{cap:Schematic-diagram-EPR}}

\end{figure}

The parametric coupling can be described conceptually by the interaction
Hamiltonian $H_{I}=i\hbar\kappa(\widehat{a}^{\dagger}\widehat{b}^{\dagger}-\widehat{a}\widehat{b})$,
which acts for a finite time $t$ corresponding to the transit time
through the nonlinear crystal. For vacuum initial states $|0,0\rangle$
this interaction generates two-mode squeezed light (\citet{CavesSchumaker1985}),
which corresponds to a quantum state in the Schrödinger picture of:\begin{eqnarray}
\left|\psi\right\rangle  & = & \sum_{{n=0}}^{\infty}c_{n}\left|n\right\rangle _{A}\left|n\right\rangle _{B}\label{eqn:twomode}\end{eqnarray}
 where $c_{n}=\tanh^{n}r/\cosh\, r$ , $r=\kappa t$, and $|n\rangle$
are number states. The parameter $r$ is called the squeezing parameter.
The expansion in terms of number states is an example of a Schmidt
decomposition, where the pure state is written with a choice of basis
that emphasizes the correlation that exists, in this case between
the photon numbers of modes $a$ and $b$. The Schmidt decomposition,
which is not unique, is a useful tool for identifying the pairs of
EPR observables (\citet{EkertKnight1995,HuangEberly1993,LawWalmsley2000}).

In our case, the EPR observables are the quadrature phase amplitudes,
as follows: \begin{eqnarray}
\hat{x} & = & \hat{x}^{A}=\widehat{a}^{\dagger}+\widehat{a},\nonumber \\
\hat{p} & = & \widehat{Y}^{A}=i\left(\widehat{a}^{\dagger}-\widehat{a}\right),\nonumber \\
\hat{x}^{B} & = & \hat{x}^{B}=\widehat{b}^{\dagger}+\widehat{b},\nonumber \\
\hat{p}^{B} & = & \widehat{Y}^{B}=i\left(\widehat{b}^{\dagger}-\widehat{b}\right).\end{eqnarray}
 The Heisenberg uncertainty relation for the orthogonal amplitudes
is $\Delta X^{A}\Delta Y^{A}\geq1$. Operator solutions at time $t$
can be calculated directly from $H_{I}$ using the rotated Heisenberg
picture, to get\begin{eqnarray}
X^{A(B)}(t) & = & X^{A(B)}(0)\cosh\left(r\right)+X^{B(A)}(0)\sinh\left(r\right)\nonumber \\
Y^{A(B)}(t) & = & Y^{A(B)}(0)\cosh\left(r\right)-Y^{B(A)}(0)\sinh\left(r\right)\label{eq:arrayX}\end{eqnarray}
 where $X^{A(B)}(0)$, $Y^{A(B)}(0)$ are the initial {}``input''
amplitudes. As $r\rightarrow\infty$, $X^{A}=X^{B}$ and $Y^{A}=-Y^{B}$,
which implies a {}``squeezing'' of the variances of the sum and
difference quadratures, so that $\Delta^{2}(X^{A}-X^{B})<2$ and $\Delta^{2}(Y^{A}+Y^{B})<2$.
The correlation of $X^{A}$ with $X^{B}$ and the anti-correlation
of $P_{A}$ with $P_{B}$, that is the signature of the EPR paradox,
is transparent, as $r\rightarrow\infty$.

The {}``EPR'' state Eq. (\ref{eqn:twomode}) is an example of a
bipartite Gaussian state, a state whose Wigner function has a Gaussian
form\begin{equation}
W(\mathbf{x})=\frac{1}{(2\pi)^{2}\sqrt{|\mathbf{C}|}}exp[-\frac{1}{2}(\mathbf{x}-\mu)^{T}\mathbf{C}^{-1}(\mathbf{x}-\mu)]\label{eq:gaus}\end{equation}
 where $\mathbf{x}=(x_{1},...,x_{4})\equiv(x,p,x^{B},p^{B})$ and
we define the mean $\mathbf{\mathbf{\mathbf{\mathbf{\mathbf{\mu}}}}}=\langle\mathbf{x}\rangle$
and the covariance matrix $\mathbf{C}$, such that $C_{ij}=\langle\hat{x}_{{i}},\hat{x}_{{j}}\rangle=\langle x_{i},x_{j}\rangle$,
$\langle v,w\rangle=\langle vw\rangle-\langle v\rangle\langle w\rangle$.
We note the operator moments of the $\hat{x}_{{i}}$ correspond directly
to the corresponding c-number moments. The state (\ref{eqn:twomode})
yields $\mu=0$ and covariance elements $C_{ii}=\Delta^{2}x_{i}=\cosh\left(2r\right)$,
$C_{13}=\langle x,x^{B}\rangle=-C_{24}=-\langle p,p^{B}\rangle=\sinh\left(2r\right)$.

We apply the linear EPR criterion of Section \ref{linearestimate}.
For the Gaussian states, in fact the \emph{best} linear estimate $x_{est}$
for $x$, given $x^{B}$, and the \emph{minimum} inference variance
$\Delta_{inf}^{2}x$ correspond to the \emph{mean} and \emph{variance}
of the appropriate conditionals, $P(x|x^{B})$ (similarly for $p$).
This mean and variance are given as in Section \ref{linearestimate}.
The two-mode squeezed state predicts, with $g=g'=\tanh\left(2r\right)$,
\begin{equation}
\Delta_{inf}^{2}x=\Delta_{inf}^{2}p=1/\cosh\left(2r\right)\,\,.\end{equation}
 Here $x=X^{A}$ is correlated with $X^{B}$, and $p=Y^{A}$ is anti-correlated
with $Y^{B}$. EPR correlations are predicted for all nonzero values
of the squeeze parameter $r$, with maximum correlations at infinite
$r$.

Further proposals for the EPR paradox that use the linear criterion,
Eq. (\ref{eq:1989crit}), have been put forward by \citet{TaraAgarwal1994}.
\citet{GiovanettiMancini2001} have presenting an exciting scheme
for demonstrating the EPR paradox for massive objects using radiation
pressure acting on an oscillating mirror.

\subsection{Measurement techniques }

Quadrature phase amplitudes can be measured using homodyne detection
techniques developed for the detection of squeezed light fields. In
the experimental proposal of \citet{DrummondReid1990}, carried out
by \citet{OuPereira1992}, an intracavity nondegenerate downconversion
scheme was used. Here the output modes are multi-mode propagating
quantum fields, which must be treated using quantum input-output theory
(\citet{CollettGardiner1984,GardinerZoller2000,DrummondFicek2004}).
Single time-domain modes are obtained through spectral filtering of
the photo-current. These behave effectively as described in the simple
model given above, together with corrections for cavity detuning and
nonlinearity that are negligible near resonance, and not too close
to the critical threshold (\citet{DechoumDrummond2004}).

At each location $A$ or $B$, a phase-sensitive, balanced homodyne
detector is used to detect the cavity output fields, as depicted in
Fig. \ref{cap:Schematic-diagram-EPR}. Here the field $\widehat{a}$
is combined (using a beam splitter) with a very intense {}``local
oscillator'' field, modeled classically by the amplitude $E$, and
a relative phase shift $\theta$, introduced to create in the detector
arms the fields $\widehat{a}_{\pm}=(\widehat{a}\pm Ee^{i\theta})/\sqrt{2}$
. Each field is detected by a photodetector, so that the photocurrent
$i_{\pm}^{A}$ is proportional to the incident field intensity $\widehat{a}_{\pm}^{\dagger}\widehat{a}_{\pm}$.
The difference photocurrent $i_{D}^{A}=i_{X}^{A}-i_{Y}^{A}$ gives
a reading which is proportional to the quadrature amplitude $X_{\theta}^{A}$,
\begin{equation}
i_{D}^{A}\propto E\hat{x}_{\theta}^{A}=E(\widehat{a}^{\dagger}e^{i\theta}+\widehat{a}e^{-i\theta})\,\,.\end{equation}
 The choice $\theta=0$ gives a measurement of $X^{A}$, while $\theta=\pi/2$
gives a measurement of $Y^{A}$. The fluctuation in the difference
current is, according to the quantum theory of detection, directly
proportional to the fluctuation of the field quadrature: thus, $\Delta^{2}i_{D}^{A}$
gives a measure proportional to the variance $\Delta^{2}X_{\theta}^{A}$.
A single frequency component of the current must be selected using
Fourier analysis in a time-window of duration $\Delta t$, which for
causality should be less than the propagation time, $L/c$.

A difference photocurrent $i_{D}^{B}$ defined similarly with respect
to the detectors and fields at $B$, gives a measure of $\hat{x}_{\phi}^{B}=\widehat{b}^{\dagger}e^{i\phi}+\widehat{b}e^{-i\phi}$.
The fluctuations in $X_{\theta}^{A}-gX_{\phi}^{B}$ are proportional
to those of the difference current $i_{D}^{A}-gi_{D}^{B}$ where $g=g^{B}/g^{A}$,
and $g^{I}$ indicates any amplification of the current $i^{I}$ before
subtraction of the currents. The variance $\Delta^{2}(i_{D}^{A}-gi_{D}^{B})$
is then proportional to the variance $\Delta^{2}(X_{\theta}^{A}-gX_{\phi}^{B})$,
so that \begin{equation}
\Delta^{2}(i_{D}^{A}-gi_{D}^{B})\propto\Delta^{2}(X_{\theta}^{A}-gX_{\phi}^{B})\,\,.\label{eq:condvarhd}\end{equation}
 In this way the $\Delta_{inf}^{2}$ of Eq. (\ref{eqn:eprcritt})
can be measured. A causal experiment can be analyzed using a time-dependent
local oscillator (\citet{Drummond1990}).

\subsection{Effects of loss and imperfect detectors}

Crucial to the validity of the EPR experiment is the accurate calibration
of the correlation relative to the vacuum limit. In optical experiments,
this limit is the vacuum noise level as defined within quantum theory.
This is represented as $1$ in the right-hand side of the criteria
in Eqs. (\ref{eqn:eprcritt}) and (\ref{eq:1989crit}).

The standard procedure for determining the vacuum noise level in the
case of quadrature measurements is to replace the correlated state
of the input field $\widehat{a}$ at $A$ with a vacuum state $|0\rangle.$
This amounts to removing the two-mode squeezed vacuum field that is
incident on the beam-splitter at location $A$ in Fig. \ref{cap:Schematic-diagram-EPR},
and measuring only the fluctuation of the current at $A$. The difference
photocurrent $i_{D}^{A}$ is then proportional to the vacuum amplitude
and the variance $\Delta^{2}i_{D}^{A}$ is calibrated to be $1$.

To provide a simple but accurate model of detection inefficiencies,
we consider an imaginary beam splitter (Fig. \ref{cap:Schematic-diagram-EPR})
placed before the photodetector at each location $A$ and $B$, so
that the detected fields $\widehat{a}$ at $A$ and $\widehat{b}$
at $B$ are the combinations $\widehat{a}=\sqrt{\eta_{A}}\widehat{a}_{0}+\sqrt{1-\eta_{A}}\widehat{a}_{vac}$
and $\widehat{b}=\sqrt{\eta_{B}}\widehat{b}_{0}+\sqrt{1-\eta_{B}}\widehat{b}_{vac}$
. Here $\widehat{a}_{vac}$ and $\widehat{b}_{vac}$ represent uncorrelated
vacuum mode inputs, $\widehat{a}_{0}$ and $\widehat{b}_{0}$ are
the original fields and $\eta_{A/B}$ gives the fractional homodyne
efficiency due to optical transmission, mode-matching and photo-detector
losses at $A$ and $B$ respectively. Details of the modeling of the
detection losses were also discussed by \citet{OuPereira1992b}. Since
the loss model is linear, the final state, although no longer pure,
is Gaussian, Eq. (\ref{eq:gaus}). Thus results concerning necessary
and sufficient conditions for entanglement/ EPR that apply to Gaussian
states remain useful. This model for loss has been experimentally
tested by \citet{BowenSchnabel2003}.

The final EPR product where the original fields are given by the two-mode
squeezed state, Eq. (\ref{eqn:twomode}), is \begin{equation}
\Delta_{inf}X^{A}\Delta_{inf}Y^{A}=1-\eta_{A}\frac{[\cosh(2r)-1][2\eta_{B}-1]}{[1-\eta_{B}+\eta_{B}\cosh(2r)]}\label{eq:infpar}\end{equation}
 We note the enhanced sensitivity to $\eta_{B}$ as compared to the
loss $\eta_{A}$ at the {}``inferred'' system $A$. It is the loss
$\eta_{B}$ at the {}``steering'' system $B$ that determines whether
the EPR paradox exists. The EPR paradox criterion (\ref{eqn:eprcritt})
is satisfied for all $\eta_{B}>0.5$, provided only that $\eta_{A},r\neq0$.
On the other hand, for all $\eta_{B}\leq0.5$ it is always the case,
at least for this situation of symmetric statistical moments for fields
at $A$ and $B$, that the EPR paradox is lost: $ $$ $$\Delta_{inf}X^{A}\Delta_{inf}Y^{A}\geq1$
(regardless of $\eta_{A}$ or $r$).

\begin{figure}
\includegraphics[width=8cm]{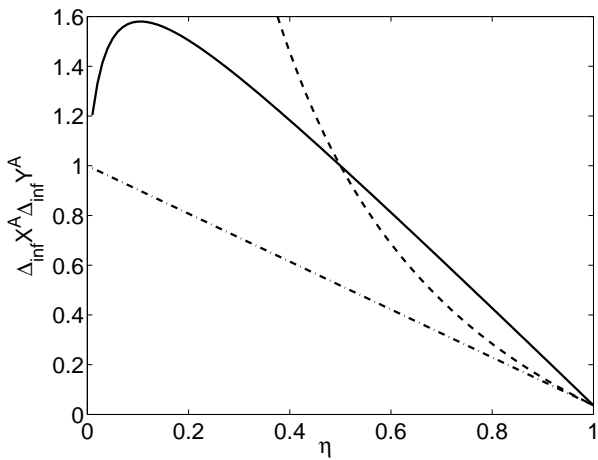}

\caption{Effect of detector efficiencies $\eta_{A}$ and $\eta_{B}$ on the
EPR paradox. Plot is $\mathcal{E}=\Delta_{inf}X^{A}\Delta_{inf}Y^{A}$
for a two-mode squeezed state with $r=2$: $\eta_{A}=\eta_{B}=\eta$
(solid line); fixed $\eta_{A}=1$ but varying $\eta=\eta_{B}$ (dashed
line); fixed $\eta_{B}=1$ but varying $\eta=\eta_{A}$ (dashed-dotted
line). The EPR paradox is sensitive to the losses $\eta_{B}$ of the
{}``steering'' system $B$, but insensitive to $\eta_{A}$, those
of the {}``inferred'' system $A$. No paradox is possible for $\eta_{B}\leq0.5$,
regardless of $\eta_{A}$, but a paradox is always possible with $\eta_{B}>0.5$,
provided only $\eta_{A}>0$.}

\label{InferredVarianceasym} 
\end{figure}

The inherently asymmetric nature of the EPR criterion is evident from
the hump in the graph of Fig. \ref{InferredVarianceasym}. This is
a measure of the error when an observer at $B$ ({}``Bob'') attempts
to infer the results of measurements that might be performed (by {}``Alice'')
at $A$. The EPR criterion reflects an absolute measure of this error
relative to the quantum noise level of field $A$ only. Loss destroys
the correlation between the signals at $A$ and $B$ so that when
loss is dominant, Bob cannot reduce the inference variance below the
fluctuation level $\Delta^{2}X^{A}$ of Alice's signal. By contrast,
calculation using the criterion of \citet{DuanGiedke2000} indicates
entanglement to be preserved for arbitrary $\eta$ (Section VII)\textbf{.}

The effect of decoherence on entanglement is a topic of current interest
(\citet{EberlyYu2007}). Disentanglement in a finite time or `entanglement
sudden death' has been reported by \citet{YuEberly2004} for entangled
qubits independently coupled to reservoirs that model an external
environment. By comparison, the continuous variable entanglement is
remarkably robust with respect to efficiency $\eta$. The death of
EPR-entanglement at $\eta=0.5$ is a different story, and applies
generally to Gaussian states that have symmetry with respect to phase
and interchange of $A$ and $B$.

A fundamental difference between the continuous-variable EPR experiments
and the experiments proposed by Bohm and Bell is the treatment of
events in which no photon is detected. These null events give rise
to loopholes in the photon-counting Bell experiments to date, as they
require fair-sampling assumptions. In continuous-variable measurements,
events where a photon is not detected simply correspond to the outcome
of zero photon number$ $ $\widehat{a}_{\pm}^{\dagger}\widehat{a}_{\pm}$,
so that $X_{\theta}^{A}=0$. These events are therefore automatically
included in the measure $\mathcal{E}$ of EPR%
\footnote{There is however the assumption that the experimental measurement
is faithfully described by the operators we assign to it. Thus one
may claim there is a loophole due to the model of loss. \citet{SkwaraKampermann2007}
discuss this point, of how to account for an arbitrary cause of lost
photons, in relation to entanglement. %
}.

Our calculation based on the symmetric two-mode squeezed state reveals
that efficiencies of $\eta>0.5$ are required to violate an EPR inequality.
This is more easily achieved than the stringent efficiency criteria
of \citet{ClauserShimony1978} for a Bell inequality violation. It
is also lower than the threshold for a spin EPR paradox (Section IV.B).
To help matters further, homodyne detection is more efficient than
single-photon detection. Recent experiments obtain overall efficiencies
of $\eta>0.98$ for quadrature detection (\citet{ZhangGoh2003,SuzukiYonezawa2006}),
owing to the high efficiencies possible when operating silicon photo-diodes
in a continuous mode.

\section{EPR, entanglement and Bell criteria}

In this Colloquium, we have understood a {}``demonstration of the
EPR paradox'' to be a procedure that closely follows the original
EPR gedanken experiment. Most generally, the EPR paradox is demonstrated
when one can confirm the inconsistency between local realism and the
completeness of quantum mechanics, since this was the underlying EPR
objective.

We point out in this Section that the inconsistency can be shown in
more ways than one. There are \emph{many uncertainty relations} or
constraints placed on the statistics of a quantum state, and for each
such relation there is an EPR criterion. This has been discussed for
the case of entanglement by \citet{Guhne2004}, and for EPR by \citet{CavalcantiReid2007}.
It is thus possible to establish a whole set of criteria that are
sufficient, but may not be necessary, to demonstrate an EPR paradox.

\subsection{{}``Steering'' }

The demonstration of an EPR paradox is a nice way to confirm the nonlocal
effect of Schrödinger's {}``steering'', a reduction of the wave-packet
at a distance (\citet{WisemanJones2007}).

An important simplifying aspect of the original EPR paradox is the
asymmetric application of local realism to imply elements of reality
for \emph{one} system, the {}``inferred'' or {}``steered'' system.
Within this constraint, we may generalize the EPR paradox, by applying
local realism to \emph{all possible} measurements, and testing for
consistency of \emph{all} the elements of reality for $A$ with a
quantum state. One may apply (\citet{CavalcantiJones2008}) the arguments
of Section IV and the approach of \citet{WisemanJones2007} to deduce
the following condition for such consistency: \begin{equation}
P(x_{\theta}^{A},x_{\phi}^{B})=\int_{\lambda}d\lambda P(\lambda)P_{Q}(x_{\theta}^{A}|\lambda)P(x_{\phi}^{B}|\lambda).\label{eqn:eprasymmetric}\end{equation}
 Here, notation is as for Eqs. (\ref{eqn:sepprob}) and (\ref{eqn:bellsep1}),
so that $P(x_{\theta}^{A},x_{\phi}^{B})$ is the joint probability
for results $x_{\theta}^{A}$ and $x_{\phi}^{B}$ of measurements
performed at $A$ and $B$ respectively, these measurements being
parametrized by $\theta$ and $\phi$. The $\lambda$ is a discrete
or continuous index, symbolizing hidden variable or quantum states,
so that $P_{Q}(x_{\theta}^{A}|\lambda)$ and $P(x_{\phi}^{B}|\lambda)$
are both probabilities for outcomes given a fixed $\lambda$. Here
as in Eq. (\ref{eqn:sepprob}), $P_{Q}^{A}(x_{\theta}^{A}|\lambda)=\langle x_{\theta}^{A}|\rho_{\lambda}|x_{\theta}^{A}\rangle$
for some quantum state $\rho_{\lambda}$, so that this probability
satisfies all quantum uncertainty relations and constraints. There
is no such restriction on $P^{B}(x_{\phi}^{B}|\lambda)$.

Eq. (\ref{eqn:eprasymmetric}) has been derived recently by \citet{WisemanJones2007},
and its failure defined as a condition to demonstrate {}``steering''.
These authors point out that Eq. (\ref{eqn:eprasymmetric}) is the
intermediate form of Eq. (\ref{eqn:sepprob}) to prove entanglement,
and Eq. (\ref{eqn:bellsep1}) used to prove failure of Bell's local
hidden variables. The failure of (\ref{eqn:eprasymmetric}) may be
considered an EPR paradox in a \emph{generalized} sense. The EPR paradox
as we define it, which simply considers a subset of measurements,
is a special case of {}``steering''.

These authors also show that for quadrature phase amplitude measurements
on bipartite Gaussian states, Eq. (\ref{eqn:eprasymmetric}) fails
when, and only when, the EPR criterion Eq. (\ref{eqn:eprcritt}) (namely
$\Delta_{inf}x\Delta_{inf}p<1$) is satisfied. This ensures that this
EPR criterion is necessary and sufficient for the EPR paradox in this
case.

\subsection{Symmetric EPR paradox }

One can extend the EPR argument further, to consider not only the
elements of reality inferred on $A$ by $B$, but those inferred on
$B$ by $A$. It has been discussed by \citet{Reid2004} that this
symmetric application implies the existence of a set of shared {}``elements
of reality'', which we designate by $\lambda$, and for which Eq.
(\ref{eqn:bellsep1}) holds. This can be seen by applying the reasoning
of the previous section to derive sets of elements of reality $\lambda_{A/B}$
for each of $A$ and $B$ (respectively), that can be then shared
to form a complete set $\left\{ \lambda_{A},\lambda_{B}\right\} $.
Explicitly, we can substitute $P(x_{\phi}^{B}|\lambda_{A})=\sum_{\lambda_{B}}P(x_{\phi}^{B}|\lambda_{B})P(\lambda_{B}|\lambda_{A})$
into (\ref{eqn:eprasymmetric}) to get (\ref{eqn:bellsep1}). Thus,
EPR's local realism can in principle be extrapolated to that of Bell's,
as defined by (\ref{eqn:bellsep1}).

Where we violate the condition (\ref{eqn:sepprob}) for separability,
to demonstrate entanglement, it is \emph{necessarily} the case that
the parameters $\lambda$ for each localized system cannot be represented
as a quantum state. In this way, the demonstration of entanglement,
for sufficient spatial separations, gives inconsistency of Bell's
local realism with completeness of quantum mechanics, and we provide
an explicit link between entanglement and the EPR paradox.

\subsection{EPR as a special type of entanglement}

While generalizations of the paradox have been presented, we propose
to reserve the title {}``EPR paradox'' for those experiments that
\emph{minimally} extend the original EPR argument, so that criteria
given in Section IV are satisfied. It is useful to distinguish the
entanglement that gives you an EPR paradox - we will define this to
be {}``EPR-entanglement'' - as a \emph{special} form of entanglement.
The EPR-entanglement is a measure of the ability of one observer,
Bob, to gain information about another, Alice. This is a crucial and
useful feature of many applications (Section X).

Entanglement itself is not enough to imply the strong correlation
needed for an EPR paradox. As shown by \citet{BowenSchnabel2003},
where losses that cause mixing of a pure state are relevant, it is
possible to confirm entanglement where an EPR paradox criterion cannot
be satisfied (Section VII). That this is possible is understood when
we realize that the EPR paradox criterion demands failure of Eq. (\ref{eqn:eprasymmetric}),
whereas entanglement requires only failure of the weaker condition
Eq. (\ref{eqn:sepprob}). The observation of the EPR paradox is a
stronger, more direct demonstration of the nonlocality of quantum
mechanics than is entanglement; but requires greater experimental
effort.

That an EPR paradox implies entanglement is most readily seen by noting
that a separable (non-entangled) source, as given by Eq. (\ref{eqn:sep}),
represents a local realistic description in which the localized systems
$A$ and $B$ are described as \emph{quantum} states $\widehat{\rho}_{\lambda}^{A/B}$.
Recall, the EPR paradox is a situation where compatibility with local
realism would imply the localized states \emph{not} to be quantum
states. We see then that a separable state cannot give an EPR paradox.
Explicit proofs have been presented by \citet{Reid2004}, \citet{MallonReid2008}
and, for tripartite situations, \citet{OlsenBradley2006}.

The EPR criterion in the case of continuous variable measurements
is written, from (\ref{eq:1989crit}) \begin{equation}
\mathcal{E}=\Delta\left(x-gx^{B}\right)\Delta\left(p+g'p^{B}\right)<1\,\,.\label{eq:EPRinequality}\end{equation}
 where $g$ and $g'$ are adjustable and arbitrary scaling parameters
that would ideally minimise $\mathcal{E}$. The experimental confirmation
of this inequality would give confirmation of quantum inseparability
\emph{on demand}, without postselection of data. This was first carried
out experimentally by \citet{OuPereira1992}.

Further criteria sufficient to prove entanglement for continuous variable
measurements were presented by \citet{DuanLukin2001} and \citet{Simon2000},
who adapted the PPT criterion of \citet{Peres1996}. These criteria
were derived to imply inseparability (entanglement) rather than the
EPR paradox itself and represent a less stringent requirement of correlation.
The criterion of \citet{DuanGiedke2000}, which gives entanglement
when \begin{equation}
D=[\Delta^{2}(x-x^{B})+\Delta^{2}(p+p^{B})]/4<1,\label{eq:duan}\end{equation}
 has been used extensively to experimentally confirm continuous variable
entanglement (refer to references of Section XI). The criterion is
both a necessary and sufficient measure of entanglement for the important
practical case of bipartite symmetric Gaussian states.

We note we achieve the correlation needed for the EPR paradox, once
$D<0.5$. This becomes transparent upon noticing that $xy\leq(x^{2}+y^{2})/2$,
and so always $\Delta(x-x^{B})\Delta(p-p^{B})\leq2D$. Thus, when
we observe $D<0.5$, we know $\Delta(x-x^{B})\Delta(p+p^{B})\leq1$,
which is the EPR criterion (\ref{eq:EPRinequality}) for $g=g'=1$.
The result also follows directly from (\ref{eq:duaneprhalfI}), which
gives, on putting $g=g'=1$, \begin{equation}
D=[\Delta^{2}(x-x^{B})+\Delta^{2}(p+p^{B})]/4<0.5\label{eq:duaneprhalf}\end{equation}
 as sufficient to confirm the correlation of the EPR paradox. We note
that this criterion, though sufficient, is not necessary for the EPR
paradox. The EPR criterion (\ref{eq:EPRinequality}) is more powerful,
being necessary and sufficient for the case of quadrature phase measurements
on Gaussian states, and can be used as a \emph{measure} of the degree
of EPR paradox. The usefulness of criterion (\ref{eq:duaneprhalfI})
is that many experiments have reported data for it. From this we can
infer an upper bound for the conditional variance product, since we
know that $ $$\mathcal{E}\leq2D$.

Recent work explores measures of entanglement that might be useful
for non-Gaussian and tri-partite states. Entanglement of formation
(\citet{BennettDiVincenzo1996}) is a necessary and sufficient condition
for all entangled states, and has been measured for symmetric Gaussian
states, as outlined by \citet{GiedkeWolf2003} and performed by \citet{JosseDantan2004a}
and \citet{GlocklAndersen2004}. There has been further work (\citet{Guhne2004,AgarwalBiswas2005,ShchukinVogel2005,HilleryZubairy2006,GuhneLutkenhaus2006})
although little that focuses directly on the EPR paradox. Inseparability
and EPR criteria have been considered however for tripartite systems
(\citet{AokiTakei2003,JingZhang2003,VanLoockFurusawa2003,BradleyOlsen2005,VillarMartinelli2006}).

\subsection{EPR and Bell's nonlocality}

A violation of a Bell inequality gives a stronger conclusion than
can be drawn from a demonstration of the EPR paradox alone, but is
more difficult to achieve experimentally. The \emph{predictions} of
quantum mechanics and local hidden variable theories are shown to
be incompatible in Bell's work. This is not shown by the EPR paradox.

The continuous variable experiments discussed in Sections VI and VII
are excellent examples of this difference. It is well-known (\citet{Bell1988})
that a local hidden variable theory, derived from the Wigner function,
exists to explain all outcomes of these continuous variable EPR measurements.
The Wigner function c-numbers take the role of position and momentum
hidden variables. For these Gaussian squeezed states the Wigner function
is positive and gives the probability distribution for the hidden
variables. Hence, for this type of state, measuring $x$ and $p$
will not violate a Bell inequality.

If the states generated in these entangled continuous variable experiments
are \emph{sufficiently pure}, quantum mechanics predicts that it is
possible to demonstrate Bell's nonlocality for \emph{other measurements}
(\citet{GrangierPotasek1988,OliverStroud1989,PraxmeyerEnglert2005}).
This is a general result for all entangled pure states, and thus also
for EPR states (\citet{GisinPeres1992}). The violation of Bell's
inequalities for continuous variable (position/ momentum) measurements
has been predicted for only a few states, either using binned variables
(\citet{LeonhardtVaccaro1995,GilchristDeuar1998,YurkeHillery1999,MunroMilburn1998,WengerHafezi2003})
or directly using continuous multipartite moments (\citet{CavalcantiFoster2007}).
An interesting question is how the degree of inherent EPR paradox,
as measured by the conditional variances of Eq. (\ref{eq:EPRinequality}),
relates quantitatively to the Bell inequality violation available.
This has been explored in part, for the Bohm EPR paradox, by \citet{FilipGavenda2004}.

It has been shown by \citet{Werner1989} that for \emph{mixed states},
entanglement does \emph{not} guarantee that Bell's local hidden variables
will fail for some set of measurements. One can have entanglement
(inseparability) without a failure of local realism. The same holds
for EPR-entanglement. For two-qubit Werner states, violation of Bell
inequalities demands greater purity ($p_{W}>0.66$ (\citet{AcinGisin2006})
than does the EPR-Bohm paradox, which can be realized for $p_{W}>0.62$
(Section IV).

\section{Continuous-wave EPR experiments}

\subsection{Parametric oscillator experiments}

The first continuous variable test of the EPR paradox was performed
by \citet{OuPereira1992}. These optically-based EPR experiments use
local-oscillator measurements with high efficiency photo-diodes, giving
overall efficiencies of more than $80\%$, even allowing for optical
losses (\citet{OuPereira1992b,GrosshansAssche2003}). This is well
above the $50\%$ efficiency threshold required for EPR.

\begin{figure}
\includegraphics[width=7cm]{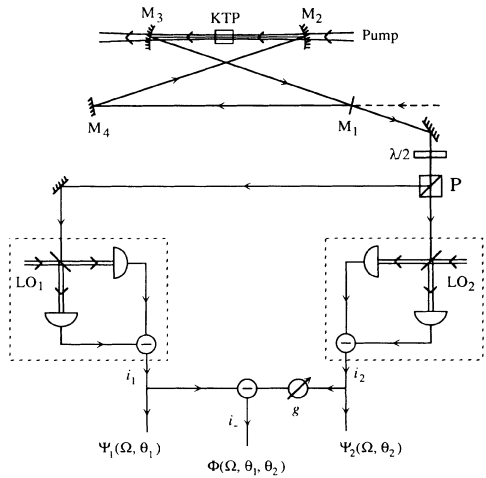}

\caption{The original EPR parametric downconversion experiment using an intracavity
nonlinear crystal and homodyne detection, following the procedure
depicted in Fig. \ref{cap:Schematic-diagram-EPR}. Figure reprinted
from \citet{OuPereira1992}, with permission.\label{cap:Parametric-downconversion-experiment}}

\end{figure}

Rather than interrogating the position and momentum of particles as
initially proposed by Einstein, Podolsky, and Rosen, analogous but
more convenient variables were used --- the amplitude and phase quadratures
of optical fields, as described in Section V. The EPR correlated fields
in the experiment of \citet{OuPereira1992} (Fig. \ref{cap:Parametric-downconversion-experiment})
were generated using a sub-threshold nondegenerate type II intra-cavity
optical parametric oscillator in a manner proposed by Reid and Drummond
(\citet{ReidDrummond1988,Reid1989,DrummondReid1990,DechoumDrummond2004}).
of a type II $\chi^{(2)}$ non-linear process in which \textit{pump}
photons at some frequency $\Omega_{\textrm{pump}}$ are converted
to pairs of correlated \textit{signal} and \textit{idler} photons
with orthogonal polarizations and frequencies satisfying $\Omega_{\textrm{signal}}+\Omega_{\textrm{idler}}=\Omega_{\textrm{pump}}$.
As discussed in Section V, these experiments utilize a spectral filtering
technique to select an output temporal mode, with a detected duration
$\Delta t$ that is typically of order $1\mu s$ or more. This issue,
combined with the restricted detector separations used to date, means
that a true, causally separated EPR experiment is yet to be carried
out, although this is certainly not impossible. In all these experiments
the entangled beams are separated and propagate into different directions,
so the only issue is the duration of the measurement. This proposal
uses cavities which are single-mode in the vicinity of each of the
resonant frequencies, so modes must be spatially separated after output
from the cavity. Another possibility is to use multiple transverse
modes together with type I (degenerate) phase-matching, as proposed
by \citet{CastelliLugiato97,OlsenDrummond2005}.

For an oscillator below threshold and at resonance, we are interested
in traveling wave modes of the output fields at frequencies $\omega_{A}$
and $\omega_{B}$. These are in an approximate two-mode squeezed state,
with the quadrature operators as given by Eq. (\ref{eq:arrayX}).
In these steady-state, continuous-wave experiments, however, the squeezing
parameter $r$ is time-independent, and given by the input-output
parametric gain $G$, such that $G=e^{2r}$. Apart from the essential
output mirror coupling, losses like absorption in the nonlinear medium
cause non-ideal behavior and reduce correlation as described in the
Section V.

Restricting ourselves to the lossless, ideal case for the moment,
we see that as the gain of the process approaches infinity ($G\rightarrow\infty$)
the quadrature operators of beams $a$ and $b$ are correlated so
that: \begin{eqnarray}
\left\langle \left(\hat{x}^{A}-\hat{x}^{B}\right)^{2}\right\rangle  & \rightarrow & 0\nonumber \\
\left\langle \left(\widehat{Y}^{A}+\widehat{Y}^{B}\right)^{2}\right\rangle  & \rightarrow & 0.\end{eqnarray}
 Therefore in this limit an amplitude quadrature measurement on beam
$a$ would provide an exact prediction of the amplitude quadrature
of beam $b$; and similarly a phase quadrature measurement on beam
$a$ would provide an exact prediction of the phase quadrature of
beam $b$. This is a demonstration of the EPR paradox in the manner
proposed in \citet{EinsteinPodolskyRosen1935}. 
 An alternative scheme is to use two independently squeezed modes
$\widehat{a}_{1},\widehat{a}_{2}$, which are combined at a $50\%$
beam-splitter so that the two outputs are $\widehat{a}_{A,B}=\left[\widehat{a}_{1}\pm i\widehat{a}_{2}\right]/\sqrt{2}$.
This leads to the same results as Eq. (\ref{eq:arrayX}), and can
be implemented if only type-I (degenerate) down-conversion is available
experimentally.

\subsection{Experimental Results}

In reality, we are restricted to the physically achievable case where
losses do exist, and the high non-linearities required for extremely
high gains are difficult to obtain. Even so, with some work at minimizing
losses and enhancing the non-linearity, it is possible to observe
the EPR paradox. Since, in general, the non-linear process is extremely
weak, one of the primary goals of an experimentalist is to find methods
to enhance it. In the experiment of \citet{OuPereira1992} the enhancement
was achieved by placing the non-linear medium inside resonant cavities
for each of the pump, signal, and idler fields. The pump field at
0.54~$\mu$m was generated by an intracavity frequency doubled Nd:YAP
laser, and the non-linear medium was a type II non-critically phase
matched KTP crystal. The signal and idler fields produced by the experiment
were analyzed in a pair of homodyne detectors. By varying the phase
of a local oscillator, the detectors could measure either the amplitude
or the phase quadrature of the field under interrogation, as described
in Section V. Strong correlations were observed between the output
photocurrents both for joint amplitude quadrature measurement, and
for joint phase quadrature measurement. To characterize whether their
experiment demonstrated the EPR paradox, and by how much, \citet{OuPereira1992}
used the EPR paradox criterion given in Eq. (\ref{eqn:eprcritt})
and Eq. (\ref{eq:1989crit}). They observed a value of $\mathcal{E}^{2}=0.70<1$,
thereby performing the first direct experimental test of the EPR paradox,
and hence demonstrating entanglement (albeit without causal separation).

\begin{figure}
\includegraphics[width=7cm]{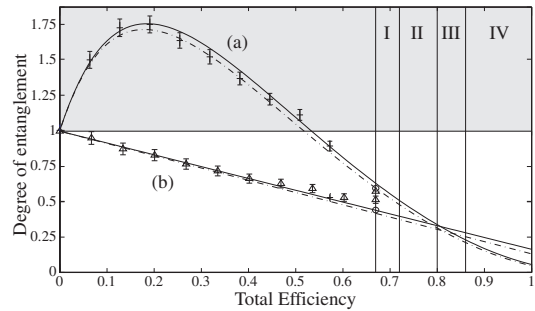}

\caption{Graph of (a) the \emph{EPR}-paradox measure $\mathcal{E}^{2}$ (Eqs.
(\ref{eqn:eprcritt}), (\ref{eq:1989crit}), (\ref{eq:EPRinequality}))
and (b) Duan \textit{et al.} (normalized) entanglement measure $D$
(Eq. (\ref{eq:duan})) vs. total efficiency $\eta.$ The dashed lines
are theoretical predictions for $ $$\mathcal{E}^{2}$ and $\mathcal{D}$.
The points are experimental data with error bars. It is more difficult
to satisfy the EPR paradox than to demonstrate entanglement. Figure
reprinted from \citet{BowenSchnabel2003}, with permission. }

\label{ExpInferredVariance} 
\end{figure}

The EPR paradox was then further tested by \citet{SilberhornLam2001,SchoriSorensen2002,BowenSchnabel2003,BowenSchnabel2004}.
Most tests were performed using optical parametric oscillators. Both
type I (\citet{BowenSchnabel2003,BowenSchnabel2004}) and type II
(\citet{OuPereira1992}) optical parametric processes, as well as
various non-linear media have been utilized. Type I processes produce
only a single squeezed field, rather than a two mode squeezed field,
so that double the resources are required in order that the two combined
beams are EPR correlated. However, such systems have significant benefits
in terms of stability and controllability. Improvements have been
made not only in the strength and stability of the interaction, but
in the frequency tunability of the output fields (\citet{SchoriSorensen2002}),
and in overall efficiency. The optimum level of EPR-paradox achieved
to date was by \citet{BowenSchnabel2003} using a pair of type I optical
parametric oscillators. Each optical parametric oscillator consisted
of a hemilithic MgO:LiNbO$_{3}$ non-linear crystal and an output
coupler. MgO:LiNbO$_{3}$ has the advantage over other non-linear
crystals of exhibiting very low levels of pump induced absorption
at the signal and idler wavelengths (\citet{FurukawaKitamura2001}
. Furthermore, the design, involving only one intracavity surface,
minimized other sources of losses, resulting in a highly efficient
process. The pump field for each optical parametric amplifier was
produced by frequency doubling an Nd:YAG laser to 532~nm. Each optical
parametric amplifier produced a single squeezed output field at 1064~nm,
with 4.1~dB of observed squeezing. These squeezed fields were interfered
on a 50/50 beam splitter, producing a two-mode squeezed state as described
in Eq. (\ref{eq:arrayX}). A degree of EPR paradox $\mathcal{E}^{2}=0.58$
was achieved. These results were verified by calibrating the loss.
The losses were experimentally varied and the results compared with
theory (Section VI), as shown in Fig. \ref{ExpInferredVariance}.
This can be improved further, as up to $9$ dB single-mode squeezing
is now possible (\citet{TakenoYukawa2007}. These experiments are
largely limited by technical issues like detector mode-matching and
control of the optical phase-shifts, which can cause unwanted mixing
of squeezed and unsqueezed quadratures.

Another technique is bright-beam entanglement above threshold, proposed
by \citet{ReidDrummond1988,ReidDrummond1989} and \citet{CastelliLugiato97}.
This was achieved recently in parametric amplifiers (\citet{VillarCruz2005,JingFeng2006,SuTan2006,VillarCassemiro2007}))
and eliminates the need for an external local oscillator. Dual-beam
second-harmonic generation can also theoretically produce EPR correlations
(\citet{LimSaffman2006}). We note that the measure $\mathcal{E}^{2}=0.58$
is to the best of our knowledge the lowest recorded result where there
has been a $ $direct measurement of an EPR paradox. A value for $ $$\mathcal{E}^{2}$
can be often be inferred from other data, either with assumptions
about symmetries (\citet{LauratCoudreau2005}), or as an upper bound,
from a measurement of the \citet{DuanGiedke2000} inseparability $D$,
since we know $\mathcal{E}\leq2D$ (Eq. (\ref{eq:duaneprhalfI}, Section
VI). Such inferred values imply measures of EPR paradox as low as
$\mathcal{E}^{2}=0.42$ (\citet{LauratCoudreau2005}, Section XI).

There has also been interest in the EPR-entanglement that can be achieved
with other variables. \citet{BowenTreps2002} obtained $\mathcal{E}^{2}=0.72$
for the EPR paradox for Stokes operators describing the field polarization.
The EPR paradox was tested for the \emph{actual} position and momentum
of single photons (\citet{FederovEfremov2004,FederovEfremov2006,GuoGuo2006})
in an important development by \citet{HowellBennink2004} to realize
an experiment more in direct analogy with original EPR. Here, however,
the exceptional value $\mathcal{E}^{2}=0.01$ was achieved using conditional
data, where detection events are only considered if two emitted photons
are simultaneously detected. The results are thus not directly applicable
to the \emph{a priori} EPR paradox. The entanglement of momentum and
position, as described in the original EPR paradox, and proposed by
\citet{CastelliLugiato97} and\citet{LugiatoGatti1997} has been achieved
using spatially entangled laser beams (\citet{WagnerJanousek2008,BoyerMarino2008}).

\section{Pulsed EPR experiments}

In the previous section we mentioned that one of the goals of an experimentalist
who aims at generating efficient entanglement is to devise techniques
by which the effective nonlinearity can be enhanced. One solution
is to place the nonlinear medium inside a cavity, as discussed above,
and another one, which will be discussed in this section, is to use
high power pump laser pulses. By using such a source the effective
interaction length can be dramatically shortened. The high finesse
cavity conditions can be relaxed or for extreme high peak power pulses,
the use of a cavity can be completely avoided. In fact a single pass
through either a highly nonlinear $\chi^{(2)}$ medium (\citet{SlusherGrangier1987,AyturKumar1990,HiranoMatsuoka1990,SmitheyBeck1992}),
or through a relatively short piece of standard glass fiber with a
$\chi^{(3)}$ nonlinear coefficient (\citet{RosenbluhShelby1991,BergmanHaus1991}),
suffices to generate quantum squeezing, which in turn can lead to
entanglement.

The limitations imposed by the cavity linewidth in the CW experiment,
such as production of entanglement in a narrow frequency band (e.g.
generation of \textquotedbl{}slow\textquotedbl{} entanglement),
are circumvented when employing a single pass pulsed configuration.
The frequency bandwidth of the quantum effects is then limited only
by the phase matching bandwidth as well as by the bandwidth of the
nonlinearity, both of which can be quite large, e.g. on the order
of some THz (\citet{SizmannLeuchs1999}). Broadband entanglement is
of particular importance for the field of quantum information science,
where for example it allows for fast communication of quantum states
by means of quantum teleportation (Section X). This may also allow
truly causal EPR experiments, which are yet to be carried out.

\subsection{Optical fiber experiment}

\begin{figure}
\includegraphics[width=7cm]{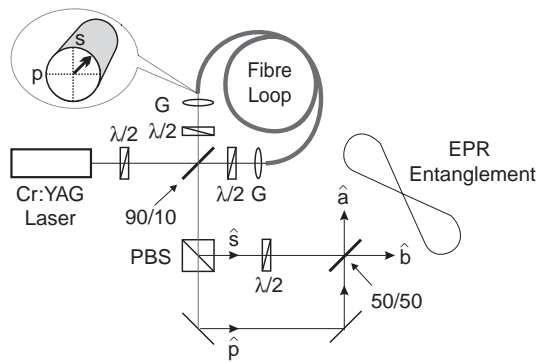}

\caption{The original demonstration of pulsed EPR entanglement. The soliton
experiment uses orthogonal polarization modes in a fiber Sagnac interferometer
and a Mach-Zehnder interferometer for fiber-birefringence compensation.
Notation: $\lambda/2$ means half-wave plate; $G$ is a gradient index
len; $50/50$ means beam splitter of 50\% reflectivity; $\hat{s}$
and $\hat{p}$ are two amplitude squeezed beams from the respective
polarization states; $\hat{a}$ and $\hat{b}$ are EPR entangled beams.
Figure reprinted from \citet{SilberhornLam2001} with permision.}

\label{Expfibre} 
\end{figure}

The first experimental realization of pulsed EPR entanglement, shown
in Fig. \ref{Expfibre} was based on the approach of mixing two squeezed
beams on a 50/50 beam splitter as outlined above for CW light. In
this experiment the two squeezed beams were generated by exploiting
the Kerr nonlinearity of silica fibers (\citet{CarterDrummond1987,RosenbluhShelby1991})
along two orthogonal polarization axes of the same polarization maintaining
fiber (\citet{SilberhornLam2001}). More precisely, the fiber was
placed inside a Sagnac interferometer to produce two amplitude squeezed
beams, which subsequently interfered at a bulk 50/50 beam splitter
(or fiber beam splitter as in \citet{NandanSabuncu2005}) to generate
two spatially separated EPR modes possessing quantum correlations
between the amplitude quadratures and the phase quadratures.

The Kerr effect is a $\chi^{(3)}$ non-linear process and is largely
equivalent to an intensity dependent refractive index. It corresponds
to a four photon mixing process where two degenerate pump photons
at frequency $\Omega$ are converted into pairs of photons (signal
and idler photons) also at frequency $\Omega$. Due to the full degeneracy
of the four-photon process, phase matching is naturally satisfied
and no external control is needed. Apart from this, optical parametric
amplification and four wave mixing are very similar (\citet{MilburnLevenson1987}).
The nonlinear susceptibility for the Kerr effect, $\chi^{(3)}$, is
very small compared to the one for optical parametric amplification,
$\chi^{(2)}$. However, as noted above, the effect is substantially
enhanced by using high peak power pulses as well as fibers resulting
in strong power confinement over the entire length of the fiber crystal.
In the experiment of \citet{SilberhornLam2001} a 16~m long polarization
maintaining fiber was used, the pulse duration was 150~fs, the repetition
rate was 163~MHz and the mean power was approximately 110~pJ. The
wavelength was the telecommunication wavelength of 1.55$\mu$m at
which the optical losses in glass are very small (0.1 dB/km) and thus
almost negligible for 16~m of fiber. Furthermore, at this wavelength
the pulses experience negative dispersion which together with the
Kerr effect enable soliton formation at a certain threshold pulse
energy, thereby ensuring a constant peak power level of the pulses
along the fiber.

The formation of solitons inside a dispersive medium is due to the
cancellation of two opposing effects - dispersion and the Kerr effect.
However, this is a classical argument and thus does not hold true
in the quantum regime. Instead, an initial coherent state is known
to change during propagation in a nonlinear medium, leading to the
formation of a squeezed state (\citet{KitagawaYamamoto1986,CarterDrummond1987,DrummondShelby1993}).
Both squeezed and entangled state solitons have been generated in
this way.

When obtaining entanglement via Kerr-induced squeezing, as opposed
to the realizations with few photons described in the previous section,
the beams involved are very bright. This fact renders the verification
procedure of proving EPR entanglement somewhat more difficult since
standard homodyne detectors cannot be used. We note that the conjugate
quadratures under interrogation of the two beams need not be detected
directly; it suffices to construct a proper linear combination of
the quadratures, e.g. $\hat{x}^{A}+\hat{x}^{B}$ and $\widehat{Y}^{A}-\widehat{Y}^{B}$.
In \citet{SilberhornLam2001} a 50/50 beam splitter (on which the
two supposedly entangled beams were interfering) followed by direct
detection of the output beams and electronic subtraction of the generated
photocurrents was used to construct the appropriate phase quadrature
combination demonstrating the phase quadrature correlations. Direct
detection of the EPR beam was employed to measure the amplitude quadrature
correlations (see also references \citet{GlocklAndersen2004,GlocklAndersen2006}).
Based on these measurements a degree of non-separability of $D=0.40$
was demonstrated (without correcting for detection losses). The symmetry
of the entangled beams allowed one to infer from this number the degree
of EPR violation, which was found to be $\mathcal{E}^{2}=0.64\pm0.08$.

The degree of entanglement as well as the purity of the EPR state
generated in this experiment were partly limited by an effect referred
to as guided acoustic wave Brillouin scattering (GAWBS) (\citet{ShelbyLevenson1985}),
which occurs unavoidably in standard fibers. This process manifests
itself through thermally excited phase noise resonances ranging in
frequency from a few megahertz up to some gigahertz and with intensities
that scales linearly with the pump power and the fiber length. The
noise is reduced by cooling the fiber (\citet{ShelbyLevenson1986}),
using intense pulses (\citet{ShelbyDrummond1990}) or by interference
of two consecutive pulses which have acquired identical phase noise
during propagation (\citet{ShirasakiHaus1992}). Recently it was suggested
that the use of certain photonic crystal fibers can reduce GAWBS (\citet{ElserAndersen2006}).
Stokes parameter entanglement has been generated exploiting the Kerr
effect in fibers using a pulsed pump source (\citet{GlocklHeersink2003}).
A recent experiment (\citet{HuntingtonMilford2005}) has shown that
adjacent sideband modes (with respect to the optical carrier) of a
single squeezed beam possess quadrature entanglement. However in both
experiments the EPR inequality was not violated, partly due to the
lack of quantum correlations and partly due to the extreme degree
of excess noise produced from the above mentioned scattering effects.

\subsection{Parametric amplifier experiment}

An alternative approach, which does not involve GAWBS, is the use
of pulsed down-conversion. Here one can either combine two squeezed
pulses from a degenerate down-conversion process, or else directly
generate correlated pulses using non-degenerate down-conversion. In
these experiments, the main limitations are dispersion (\citet{RaymerDrummond1991})
and absorption in the nonlinear medium. \citet{WengerOurjoumtsev2005}
produced pulsed EPR beams, using a traveling-wave optical parametric
amplifier pumped at 423 nm by a frequency doubled pulsed Ti:Sapphire
laser beam. Due to the high peak powers of the frequency doubled pulses
as well as the particular choice of a highly non-linear optical material
(KNBO$_{3}$), the use of a cavity was circumvented despite the fact
that a very thin (100 $\mu m$) crystal was employed. A thin crystal
was chosen in order to enable broadband phase matching, thus avoiding
group-velocity mismatch. The output of the parametric amplifier was
then a pulsed two-mode squeezed vacuum state with a pulse duration
of 150 fs and a repetition rate of 780 kHz.

In contrast to the NOPA used by \citet{OuPereira1992}, which was
non-degenerate in polarization, the process used by Wenger \textit{et
al.} was driven in a spatially non-degenerate configuration so the
signal and idler beams were emitted in two different directions. In
this experiment the entanglement was witnessed by mixing the two EPR
beams with a relative phase shift of $\phi$ at a 50/50 beam splitter
and then monitoring one output using a homodyne detector. Setting
$\phi=0$ and $\phi=\pi$, the combinations $\hat{x}^{A}+\hat{x}^{B}$
and $\widehat{Y}^{A}-\widehat{Y}^{B}$ were constructed. They measured
a non-separability of $D=0.7$ (without correcting for detector losses).
Furthermore the noise of the individual EPR beams were measured and
all entries of the covariance matrix were estimated (assuming no inter-
and intra-correlations).

Without correcting for detector inefficiencies we deduce that the
EPR paradox was not demonstrated in this experiment since the product
of the conditional variances amounts to $\mathcal{E}^{2}=1.06$. However,
by correcting for detector losses as done in the paper by Wenger \textit{et
al.}, the EPR paradox was indeed achieved since in this case the EPR-product
is $\mathcal{E}^{2}=0.83$, although causal separation was not demonstrated.
A degenerate waveguide technique, together with a beam-splitter, was
recently used to demonstrate pulsed entanglement using a traveling
wave OPA (\citet{ZhangFuruta2007}).

A distinct difference between the two pulsed EPR experiments, apart
from the non-linearity used, is the method by which the data processing
was carried out. In the experiment by \citet{SilberhornLam2001} ,
measurements were performed in the frequency domain similar to the
previously discussed CW experiments: The quantum noise properties
were characterized at a specific Fourier component within a narrow
frequency band, typically in the range 100-300 kHz. The frequency
bandwidth of the detection system was too small to resolve successive
pulses, which arrived at the detector with a frequency of 163 MHz.
In the experiment of Wenger \textit{et al.,} however, the repetition
rate was much lower (780kHz), which facilitated the detection stage
and consequently allowed for temporally-resolved measurements around
DC (\citet{SmitheyBeck1992,SmitheyBeck1993}).

\section{Spin EPR and atoms }

Experimental realizations of the paradox with massive particles are
important, both due to their closeness in spirit with the original
EPR proposal, and because such massive entities could reasonably be
considered more closely bound to the concept of local realism than
fields. To date, experimental tests of the EPR paradox with massive
particles have been limited to situations of small spatial separation.
However, the technology required to generate, manipulate, and interrogate
non-classical states of massive systems has undergone rapid development
over the past decade. These often involve spin-equivalent versions
of the EPR paradox with spin quantum numbers much larger than one
half. A spin-one (four-particle) Bell inequality violation of a type
predicted by \citet{Drummond1983} was observed experimentally by
\citet{HowellLamasLinares2002}. Criteria for observing a spin EPR
paradox and the experimental test of \citet{BowenSchnabel2002} have
been discussed in Section IV.B.


Many theoretical proposals and experimental techniques to entangle
pairs of atoms and atomic ensembles have been developed (\citet{CiracZoller1997}).
The core technologies involved range from single neutral atoms trapped
in high-$Q$ optical microresonators and manipulated with optical
pulses (\citet{Kimble1998,McKeeverBuck2003}), to multiple ions trapped
in magnetic traps with interaction achieved through vibrational modes,
to optically dense ensembles of atoms (\citet{Polzik1999,JulsgaardKozhekin2001,JulsgaardSherson2004,KuzmichMandel2000}).

Future experiments on ultra-cold atoms may involve direct entanglement
of the atomic position. Possible experimental systems were recently
analyzed by \citet{FederovEfremov2006}, for pairs of massive or massless
particles. Another approach for EPR measurements is to use correlated
atom-laser beams generated from molecular dissociation (\citet{KheruntsyanOlsen2005}).
This proposal involves macroscopic numbers of massive particles, together
with superpositions of different spatial mass-distributions. Entanglement
of this type therefore could test the unification of quantum theory
with gravity.

Here we focus on experiments based on atomic ensembles, which have
shown the most promise for tests of the EPR paradox. In these, a weak
atom-light interaction is used to generate a coherent excitation of
the spin state of a large number of atoms within the ensemble.Through
appropriate optical manipulation, both squeezing and entanglement
of this collective macroscopic spin state have been demonstrated (\citet{GeremiaStockton2004,KuzmichMoelmer1997,KuzmichMandel2000,HaldSorensen1999}),
as well as entanglement of spatially separated atomic ensembles (\citet{JulsgaardSherson2004,ChouRiedmatten2005,ChaneliereMatsukevich2005,MatsukevichChaneliere2006}).

Decoherence is a critical factor which limits the ability to generate
squeezing and entanglement in atomic systems. One might expect that
since spin-squeezed and entangled atomic ensembles contain a large
number $N$ of atoms, the decoherence rate of such systems would scale
as $N\gamma$ where $\gamma$ is the single atom decay rate. Indeed,
this is the case for other multi-particle entangled states such as
Greenberger-Horne-Zeilinger entanglement (\citet{GreenbergerHorne1989}).
However, a critical feature of these collective spin states is that
excitation due to interaction with light is distributed symmetrically
amongst all of the atoms. This has the consequence that the system
is robust to decay (or loss) of single atoms. Consequently, the decoherence
rate has no dependence on $N$ and is equal to the single photon decay
rate $\gamma$ (\citet{Lukin2003}). Several experimental techniques
have been developed to further reduce the decoherence rate. These
include the use of buffer gases (\citet{PhillipsFleischhauer2001})
and paraffin coatings (\citet{JulsgaardKozhekin2001}) in room temperature
vapor cells to respectively minimize collisions between atoms and
the effect of wall collisions; and the use of cold atoms in magneto-optic
traps (\citet{GeremiaStockton2004}). These techniques have lead to
long decoherence times of the order of 1 ms for the collective spin
states.

\subsection{Transfer of optical entanglement to atomic ensembles}

The work of \citet{Polzik1999} showed that the optical entanglement
generated by a parametric oscillator, as described in Section VII
could be transferred to the collective spin state of a pair of distant
atomic ensembles. This research built on earlier work focusing on
the transfer of optical squeezing to atomic spin states (\citet{KuzmichMoelmer1997}).
In both cases, however, at least 50\% loss was introduced due to spontaneous
emission. As discussed in Section V, the EPR paradox cannot be tested
when symmetric losses that exceed $50$\%. Therefore, the proposal
of \citet{Polzik1999} is not immediately suitable for tests of the
EPR paradox. Extensions of this work have shown that by placing the
atomic ensemble within an optical resonator, the quantum state transfer
can be enhanced so that tests of the EPR paradox should be possible
(\citet{DantanPinard2003,VernacPinard2001}).

The first experimental demonstration of quantum state transfer from
the polarization state of an optical field to the collective spin
state of an atomic ensemble was performed by \citet{HaldSorensen1999}.
They demonstrated transfer of as much as -0.13 dB of squeezing to
an ensemble of $10^{9}$ cold atoms in a magneto-optic trap. The extension
of these results to pairs of spatially separated entangled ensembles
has yet to be performed experimentally.



\subsection{Conditional atom ensemble entanglement }

The other approach to experimental demonstration of collective spin
entanglement in atomic ensembles is to rely on conditioning measurements
to prepare the state (\citet{JulsgaardSherson2004,ChouRiedmatten2005}).
This approach has the advantage of not requiring any non-classical
optical resources. \citet{KuzmichMandel2000} performed an experiment
that was based on a continuous quantum non-demolition (QND) measurement
of the $z$ spin projection of a room temperature ensemble of spin-polarized
Cesium atoms in a paraffin-coated glass cell and demonstrated 5.2
dB of collective spin squeezing. A subsequent experiment along theses
lines by \citet{GeremiaStockton2004} utilized control techniques
to further enhance the generation of QND based collective spin squeezing.
The definition of collective spin in extended atomic systems of this
type is discussed in \citet{DrummondRaymer1991}.

In a major advance, collective spin entanglement was generated by
\citet{JulsgaardKozhekin2001} using techniques similar to the QND
measurements above. They interacted a pulse of light with two spatially
separated spin-polarized atomic ensembles in paraffin-coated glass
cells, and performed a nonlocal Bell measurement on the collective
spin through detection of the transmitted pulse. This conditioned
the state of the atomic ensembles into a collective entangled state
of the type required to test the EPR paradox. They report that if
utilised in a unity gain coherent state teleportation experiment,
this atomic entanglement could allow a fidelity as high as 0.55. This
corresponds to an inseparability value of D = 0.82, which is well
below 1 (indicating entanglement), but is not sufficient for a direct
test of the EPR paradox.

Recently, techniques to condition the spin state of atomic ensembles
have been developed based on the detection of stimulated Raman scattering.
These techniques have significant potential for quantum information
networks (\citet{DuanLukin2001}) and are also capable of generating
a collective entangled state of the form required to test the EPR
paradox. The experiment by \citet{KuzmichBowen2003} demonstrated
non-classical correlations between pairs of time-separated photons
emitted from a Cs ensemble in a magneto-optical trap. Through the
detection of the second photon the atomic ensemble was conditioned
into a non-classical state. The principle of the experiment by \citet{WalEisaman2003}
was the same. However, a Rb vapor cell with buffer gas was used, and
field quadratures were detected rather than single photons. This experiment
demonstrated joint-squeezing of the output fields from the ensemble,
implying the presence of collective spin squeezing within the ensemble.
Entanglement between two spatially separate ensembles has now been
demonstrated based on the same principles (\citet{ChouRiedmatten2005,MatsukevichChaneliere2006}).

\section{Application of EPR entanglement}

Entanglement is a central resource in many quantum information protocols.
A review of the continuous variable quantum information protocols
has been given by \citet{BraunseteinvanLoock2005}. In this section,
we focus on three continuous-variable quantum information protocols
that utilize shared EPR entanglement between two parties. They are
entanglement-based quantum key distribution, quantum teleportation
and entanglement swapping. We discuss the relevance of the EPR paradox
in relation to its use as a figure of merit for characterizing the
efficacy of each of these protocols.

\subsection{Entanglement-based quantum key-distribution}

In quantum key distribution (QKD), a sender, Alice, wants to communicate
with a receiver, Bob, in secrecy. They achieve this by first cooperatively
finding a method to generate a secret key that is uniquely shared
between the two of them. Once this key is successfully generated and
shared, messages can be encrypted using a {}``one-time-pad'' algorithm
and communication between them will be absolutely secure. Figure \ref{cap:Schematic-diagram-EPR}
shows that the EPR paradox can be demonstrated when Alice and Bob
get together to perform conditional variance measurements of the quadrature
amplitudes of a pair of entangled beams. The product of the conditional
variances of both quadrature amplitudes gives the degree of EPR entanglement.
Since EPR entangled beams cannot be cloned, it has been proposed by
\citet{Reid2000} and \citet{SilberhornKorolkova2002} that the sharing
of EPR entanglement between two parties can be used for QKD.

In order to use the EPR entanglement for QKD, we assume that the entanglement
generation is performed by Alice. Alice keeps one of the entangled
beams and transmits the other to Bob. It is therefore reasonable to
assume that Alice's measurements on her beam has negligible loss by
setting $\eta_{A}=1$ whilst Bob's measurements are lossy due to the
long distance transmission of entanglement with $\eta_{B}<1$. With
Alice and Bob both randomly switching their quadrature measurement
between amplitude ($X^{A}$ for Alice and $X^{B}$ for Bob) and phase
($Y^{A}$ for Alice and $Y^{B}$ for Bob), the secret key for the
cryptographic communication is obtained from the quantum fluctuations
of the EPR entanglement when there is an agreement in their chosen
quadrature.

Since the results of measurements between Alice and Bob are never
perfectly identical, Alice and Bob are required to reconcile the results
of their measurements. Conventionally, it was assumed that Bob is
required to guess Alice's measured values. The net information rate
for QKD, as suggested by \citet{CsiszarKorner1978}, is given by \begin{equation}
\Delta I=\frac{1}{2}\log_{2}\left(\frac{V_{A|E}^{X}V_{A|E}^{Y}}{V_{A|B}^{X}V_{A|B}^{Y}}\right)\end{equation}
 where $V_{A|B}^{X}=\Delta_{inf}^{2}X^{A}$ and $V_{A|B}^{Y}=\Delta_{inf}^{2}Y^{A}$
are the conditional variances defined in Section \ref{linearestimate}
for inferences made about $A$ from $B$, and where $V_{A|E}^{X,Y}$
is calculated by assuming that an eavesdropper Eve has access to all
of the quantum correlations resulting from transmission losses. When
the net information rate is positive, $\Delta I>0$, a secret key
can be generated between Alice and Bob. The conditional variance product
$V_{A|B}=\Delta_{inf}^{2}X^{A}\Delta_{inf}^{2}Y^{A}$ can be written:
\begin{equation}
V_{A|B}=\left[V_{A}^{X}-\frac{\left|\left\langle \hat{x}^{B},\hat{x}^{A}\right\rangle \right|^{2}}{V_{B}^{X}}\right]\left[V_{A}^{Y}-\frac{\left|\left\langle \widehat{Y}^{B},\widehat{Y}^{A}\right\rangle \right|^{2}}{V_{B}^{Y}}\right]\end{equation}
 Here we define $V_{A,B}^{X}=\Delta^{2}X^{A,B}$, and $V_{A,B}^{Y}=\Delta^{2}Y^{A,B}$.
We note from Fig. \ref{InferredVarianceasym} that $V_{A|B}>1$ for
$\eta_{B}<0.5$. This suggests that Alice and Bob can no longer share
EPR entanglement for larger than 3~dB transmission loss. This loss
limit is referred to as the 3 dB limit for QKD.

If on the other hand, Alice was to infer Bob's measured results, the
relevant EPR measure and net information rate are respectively given
by\begin{eqnarray}
V_{B|A} & = & \left[V_{A}^{X}-\frac{\left|\left\langle \hat{x}^{B},\hat{x}^{A}\right\rangle \right|^{2}}{V_{A}^{X}}\right]\left[V_{B}^{Y}-\frac{\left|\left\langle \widehat{Y}^{B},\widehat{Y}^{A}\right\rangle \right|^{2}}{V_{A}^{Y}}\right]\nonumber \\
\Delta I & = & \frac{1}{2}\log_{2}\left(\frac{V_{B|E}^{X}V_{B|E}^{Y}}{V_{B|A}^{X}V_{B|A}^{Y}}\right)\end{eqnarray}
 Fig. \ref{InferredVarianceasym} suggests that it is possible to
have $V_{B|A}\le1$ and $\Delta I>0$ for all values of $0<\eta_{B}<1$.
Entanglement can thus exist over long distances and the 3~dB limit
for entanglement-based QKD can be surpassed.

The advantage gained by reversing the inference, known as \textit{reverse
reconciliation}, was first recognized by \citet{GrosshansAssche2003}.
It can be simply understood as follows. When Bob and Eve both attempt
to infer the information Alice sent using their respective measurements,
a greater than $50\%$ loss where $\eta_{B}<0.5$ will give Eve an
irrecoverable information advantage over Bob since one has to assume
that Eve somehow has access to more than 50\% of the information.
In reverse reconciliation, Alice and Eve will both attempt to infer
Bob's results. Since Alice's entanglement is assumed to be lossless
($\eta_{A}=1$), she maintains her information advantage relative
to Eve, who only has partial information that is at most proportional
to transmission losses.


\subsection{Quantum Teleportation and Entanglement Swapping}

\begin{figure}[!h]
 \includegraphics[width=8cm]{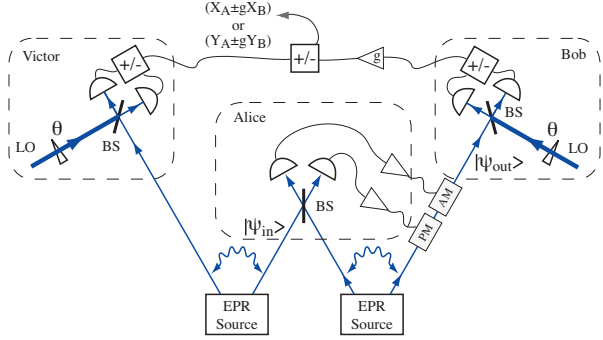}

\caption{(Color online) Schematic of quantum teleportation and entanglement
swapping. In teleportation, Alice and Bob share a pair of entangled
beams. $|\psi_{{\rm in}}\rangle$ is the input state Alice teleports
to Bob. The use of electro-optic feedforward on both the amplitude
and phase quadrature on Bob's entangled beam produces an output state
$|\psi_{{\rm out}}\rangle$ which he measures using optical homodyne
detection, as in Fig. \ref{cap:Schematic-diagram-EPR}. In entanglement
swapping, Alice and Victor also share a pair of entangled beams. Alice
uses her share of this pair as the input state $|\psi_{{\rm in}}\rangle$.
The teleportation protocol is again performed. Victor verifies the
efficacy of entanglement swapping using conditional variance measurements
of his entangled beam with Bob's teleportation output beam. The elements
are: beam splitters BS, local oscillator LO, phase shift $\theta$,
difference/ sum currents $+/-$. Semicircles are photodiodes, while
triangles show electronic gain.}

\label{EntgSwap} 
\end{figure}

Quantum teleportation is a three stage protocol that enables a sender,
Alice, to transmit a quantum state to a receiver, Bob, without a direct
quantum channel. Fig. \ref{EntgSwap} gives the schematic of the protocol.
Alice first makes simultaneous measurements of a pair of conjugate
observables of an unknown quantum state, $|\psi\rangle$, by interfering
the unknown quantum state with one of the entangled beam pairs she
shares with Bob. She then transmits both her measured results to Bob
using two classical channels. Using the other entangled beam, Bob
reconstructs the quantum state by manipulation of the other entangled
beam, using the classical information obtained from Alice. In an ideal
situation, the output state of Bob will be an exact replica of the
unknown input state sent by Alice. This form of remote communication
of quantum information using only entanglement and classical information
was proposed by \citet{BennettBrassard1993} for discrete variables.
A year later, \citet{Vaidman1994} extended this idea to allow for
continuous-variable systems, such as the teleportation of position
and momentum of a particle or the quadrature amplitudes of a laser
beam. Further work on continuous-variable quantum teleportation by
\citet{BraunsteinKimble1998} and \citet{RalphLam1998} shows that
quantum teleportation can indeed be demonstrated using finite squeezing
and entanglement.

For realistic experimental demonstration of continuous-variable quantum
teleportation, the output state cannot be identical to the teleporter
input because of the finite quantum correlations available in experimentally
produced squeezing and entanglement. A well accepted measure of teleportation
efficacy is the overlap of the wavefunction of the output state with
the original input state. The teleportation fidelity is given by $ $${\cal F}=\langle\psi_{{\rm in}}|\widehat{\rho}_{{\rm out}}|\psi_{{\rm in}}\rangle$
where $\widehat{\rho}_{{\rm out}}$ is the density operator of the
output state. Ideally, quantum teleportation can give a fidelity of
unity. For a Gaussian distribution of coherent states, with mean photon
number $\overline{n}$, the average fidelity using classical measure
and regenerate strategies is limited to ${\cal F}<(\overline{n}+1)/(2\overline{n}+1)$
(\citet{HammererWolf2005}. In the limit of large photon number, one
obtains ${\cal F}<0.5$, commonly referred to as the classical limit
for fidelity. Experiments with teleportation fidelity surpassing this
limit were demonstrated by \citet{FurusawaSorensen1998}, \citet{ZhangGoh2003}
and \citet{BowenTreps2003}. More recently \citet{GrosshansGrangier2001}
suggested that for ${\cal F}>2/3$, Bob's output state from the teleporter
is the best reconstruction of the original input. Alice, even with
the availability of perfect entanglement, cannot conspire with another
party to replicate a better copy than what Bob has reconstructed.
This average fidelity value is referred to as the no-cloning limit
for quantum teleportation. This limit has been experimentally surpassed
by \citet{TakeiYonezawa2005}.

The use of fidelity for characterizing teleportation has limitations.
Firstly, fidelity captures only the mean value behavior of the output
state relative to the input. The measure does not directly guarantee
that quantum fluctuations of the input state are faithfully replicated.
Secondly, fidelity is an input-state dependent measure. In theory,
measurements of fidelity have to be averaged over a significant region
of the quadrature amplitude phase space before the suggested bounds
are valid classical and no-cloning limits. Alternatively, \citet{RalphLam1998}
suggested that the measure of the EPR paradox can be used to characterize
quantum teleportation. The teleportation efficacy can be measured
in terms of the conditional variance measure, V, and an additional
information transfer coefficient, T, given by\begin{eqnarray}
V_{A|B} & = & \left[V_{out}^{X}-\frac{\left|\left\langle \hat{x}_{in},\hat{x}_{out}\right\rangle \right|^{2}}{V_{in}^{X}}\right]\left[V_{out}^{Y}-\frac{\left|\left\langle \widehat{Y}_{in},\widehat{Y}_{out}\right\rangle \right|^{2}}{V_{in}^{Y}}\right]\nonumber \\
T & = & \frac{{\cal R}_{{\rm out}}^{X}}{{\cal R}_{{\rm in}}^{X}}+\frac{{\cal R}_{{\rm out}}^{Y}}{{\cal R}_{{\rm in}}^{Y}}\,\,.\end{eqnarray}
 where ${\cal R}$ is the signal-to-noise variance ratio, and $X$,
$Y$ are the quadratures for the respective input and output states.
$V$ is therefore a direct measure of the correlations of quantum
fluctuations between the input and the output state. $T$, on the
other hand, measures the faithful transfer of information of both
quadrature amplitudes. Without the use of shared entanglement, it
can be shown that quantum teleportation is limited to $V\ge1$ and
$T\le1$ (\citet{RalphLam1998,BowenTreps2003}).

Unlike teleportation fidelity, it can be shown that these $T-V$ parameters
are less dependent on input states. Their direct measurements does,
however, pose some problems. Since the teleported input is invariably
destroyed by Alice's initial measurements, Bob cannot in real time
directly work out the conditional variances of his output state relative
to the destroyed input. Nevertheless, by making a suitable assumption
of the gain of the teleporter, an inferred conditional variance product
can be calculated.

The difficulty in directly measuring the conditional variance product
is resolved when we consider using a beam from another entanglement
source as the input state, as shown in Fig. \ref{EntgSwap}. The teleported
output of this entangled beam can be interrogated by the $T-V$ as
suggested. This protocol is known as \textit{entanglement swapping.}
The first continuous variable entanglement swapping experiment was
reported by \citet{TakeiYonezawa2005}.

\section{Outlook}

\begin{figure}
\includegraphics[width=8cm]{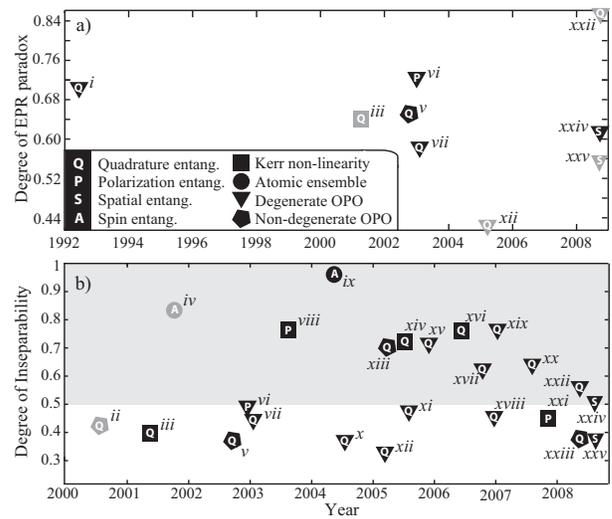} \label{EPRHist}

\caption{a) A history of experiments investigating measure of: a) $\mathcal{E}^{2}$,
the EPR paradox (Eq. (\ref{eqn:eprcritt})) and b) inseparability
$D$ (Eqs. (\ref{eq:duan}) and (\ref{eq:duaneprhalfI})), for continuous
variable measurements. Where $D<0.5$, one can infer an EPR paradox,
using $\mathcal{E}\leq2D$ (Section VI). The grey labels in (a) indicate
that $\mathcal{E}^{2}$ has not been measured directly, but is inferred
by the authors. From (b) we see that an EPR paradox could have been
inferred in other experiments as well. ($i$) \citet{OuPereira1992},
($ii$) \citet{ZhangWang2000}){[}Inferred from a variance product
measurement], ($iii$) \citet{SilberhornLam2001}, ($iv$) \citet{JulsgaardKozhekin2001},
($v$) \citet{SchoriSorensen2002}, ($vi$) \citet{BowenTreps2002},
($vii$) \citet{BowenSchnabel2003}, ($viii$) \citet{GlocklHeersink2003},
($ix$) \citet{JosseDantan2004a}, ($x$) \citet{HayasakaZhang2004},
($xi$) \citet{TakeiYonezawa2005}, ($xii$) \citet{LauratCoudreau2005},
($xiii$) \citet{WengerOurjoumtsev2005}, ($xiv$) \citet{HuntingtonMilford2005},
($xv$) \citet{VillarCruz2005}, ($xvi$) \citet{NandanSabuncu2005},
($xvii$) \citet{JingFeng2006}, (\emph{xviii}) \citet{Takei2006},
(\emph{xix}) \citet{Yoshino2007}, (\emph{xx}) \citet{ZhangFuruta2007},
(\emph{xx}i) \citet{DongHeersink2007}, (\emph{xxii)} \citet{KellerDauria2008},
(\emph{xxiii)} \citet{GrosseAssad08}, (\emph{xxi}v) \citet{WagnerJanousek2008},
(\emph{xxv}) \citet{BoyerMarino2008}. Inseparability has also been
verified using other measures, such as negativity (\citet{OurjoumtsevDantan2007}).}

\end{figure}

The Einstein-Podolsky-Rosen gedanken-experiment has been realized
through a series of important developments, both theoretical and technological.
Experiments have measured violation of the inferred Heisenberg uncertainty
principle, thus confirming EPR-entanglement. Fig. 9 summarizes the
degree of entanglement and the degree of EPR paradox achieved in continuous
variable experiments to date.

A question often arising is the utility of such measurements, given
that Bell inequality violations are a more powerful indication of
the failure of local realism. There are multiple reasons for this.
The beauty of the EPR approach is its simplicity, both from a theoretical
and a practical point of view. Bell inequalities have proved in reality
exceedingly difficult to violate. EPR measurements with quadratures
do not involve conditional state preparation or the inefficient detectors
found in most current photon-based Bell inequality experiments, and
the issue of causal separation does not look insurmountable.

The development of these techniques also represents a new technology,
with potential applications in a number of areas ranging from quantum
cryptography and ultra-precise measurements, through to innovative
new experimental demonstrations of ideas like quantum `teleportation'
- using entanglement and a classical channel for transmission of quantum
states between two locations.

Owing to Bell's theorem, Einstein \emph{et al}.'s argument for completing
quantum mechanics is sometimes viewed as a mistake. Yet there exist
alternatives to standard quantum theory which are \emph{not} ruled
out by any Bell experiments. These include spontaneous decoherence
(\citet{GhirardiRimini1986,BassiGhirardi2003}), gravitational nonlinearity
(\citet{Penrose1998,Diosi2007}), and absorber theories (\citet{Pegg1997}).
By using field-quadrature measurements and multi-particle states,
quantum theory and its alternatives can be tested for increasingly
macroscopic systems (\citet{Marshall2003}). However, an ingredient
central to the EPR argument, causal separation of measurement events,
is missing from these experiments to date. In view of this, further
EPR experiments are of considerable interest, especially with causal
separation and/or massive particles. 
\begin{acknowledgments}
We thank S. Pereira, J. H. Eberly, M. Hillery, C. Fabre, J. Kimble,
H. Wiseman, Q. He and acknowledge support from the ARC Centre of Excellence
Program and Queensland State Government. EGC acknowledges funding
through Griffith University's research fellowship scheme and H. Wiseman's
Federation Fellowship FF0458313. ULA and GL acknowledge support from
the EU project COMPAS and the Danish Research Council (FTP).

\bibliographystyle{apsrmp} \bibliographystyle{apsrmp}
\bibliography{EPR}

\end{acknowledgments}

\end{document}